\documentclass[twocolumn,usenatbib]{mnras}
\usepackage{graphicx}
\usepackage{lscape}
\usepackage{bookmark}
\usepackage{threeparttablex}
\usepackage{tabularx}
\usepackage[T1]{fontenc}
\usepackage{ae,aecompl}
\usepackage{adjustbox}
\usepackage{subcaption}
\usepackage{savesym}
\usepackage{amsmath}
\usepackage{gensymb}
\savesymbol{iint}
\usepackage{wasysym}
\usepackage{graphicx}	
\usepackage{amssymb}	
\usepackage{etoolbox}
\usepackage{threeparttable}
\usepackage{environ}
\usepackage{hyperref}
\usepackage{rotating}
\usepackage{adjustbox}

\usepackage{soul}

\title[Study of NGC 6940 using UVIT/\textit{AstroSat}]{UOCS-XII. A study of open cluster NGC 6940  using UVIT/\textit{AstroSat}: cluster properties and exotic populations}
\author[A. Panthi \& K. Vaidya]{
Anju Panthi $^{1}$\thanks{p20190413@pilani.bits-pilani.ac.in}
and Kaushar Vaidya,$^{1}$  
\\
\\
$^{1}$ Department of Physics, Birla Institue of Technology and Science, Pilani, Rajasthan-333031, India\\
}

\begin{document}
\label{firstpage}
\pagerange{\pageref{firstpage}--\pageref{lastpage}}
\maketitle

\begin{abstract}
We study an open cluster NGC 6940 using \textit{AstroSat}/UVIT data and other archival data. This is an intermediate age cluster ($\sim$ 1 Gyr), located at about 770 pc distance, harboring several exotic populations apart from normal single and binary stars. We identify members of this cluster using a machine learning algorithm, ML-MOC and identify 492 members, including 1 blue straggler star (BSS), 2 yellow straggler stars (YSS), 11 blue lurker (BL) candidates, and 2 red clump (RC) stars. The cluster shows the effect of mass segregation, with massive stars segregated the most into the cluster, followed by the equal-mass binary members and the single low mass stars. We report the presence of an extended main-sequence turn-off (eMSTO) feature in this cluster and suggest that the age spread may be a contributing factor behind it. However, the effect of stellar rotation, and the dust absorption needs to be examined more comprehensively with a larger fraction of MSTO stars. All the sixteen sources mentioned above have a counterpart in the UVIT/F169M filter. In order to characterize them, we construct multi-wavelength spectral energy distributions (SEDs) of 14 of these objects having no nearby sources within 3$\arcsec$. The BSS is successfully fitted with a single-component SED. We find that three BLs, two YSS, and one RC star have UV excess greater than 50$\%$ and successfully fit two-component SEDs having hot companions. The parameters derived from the SEDs imply that the hot companions of BLs and the RC star are low-mass and normal-mass white dwarfs, whereas the hot companions of YSS are likely to be subdwarf B (sdB) stars. We suggest that at least 6 out of 14 stars ($\sim$42 $\%$) are formed via mass transfer and/or merger pathways. 
\end{abstract}


\section{Introduction} \label{Introduction}

Star clusters serve as excellent environments for studying diverse stellar populations within their host galaxies. They consist of a uniform collection of stars that share the same distances, kinematics, metallicities, and ages, although several open clusters younger than $\sim$ 2 Gyr exhibit a feature known as an extended main-sequence turn-off (eMSTO), which may be due to the age spread within the cluster members. Star clusters, therefore, offer an ideal opportunity to investigate the evolutionary processes of both single and binary stars. The occurrence of exotic stars within star clusters is often associated with the evolution of stars in binary or multiple systems. These systems are often formed through internal binary evolution or as a result of dynamic interactions involving equal-mass binaries and single stars \citep{yang2013effects,schneider2015evolution}. Notable examples of intriguing stellar varieties found in star clusters encompass blue straggler stars (BSS), blue lurkers (BLs), yellow straggler stars (YSS), and red clump (RC) stars.
 
BSS exhibit greater brightness and blueness compared to the main sequence turn-off (MSTO), as seen on the colour-magnitude diagram (CMD). These stars were discovered by \cite{sandage1953color} in the globular cluster M3. The mechanisms behind their formation and evolution remain poorly understood due to several reasons, such as BSS formed via different mechanisms are indistinguishable photometrically and high-resolution spectroscopic observations to detect abundance variations are demanding on time and resources. BSS appear to have acquired additional mass through unconventional processes, leading to their rejuvenation. Several explanations have been proposed for their formation, but three mechanisms are widely acknowledged in the literature. Firstly, stellar collisions involving two or more stars in dense environments \citep{hills1976stellar} are considered a possible mechanism. Secondly, mass transfer (MT) through Roche lobe overflow in a binary system \citep{mccrea1964extended} is another commonly accepted explanation. Lastly, MT leading to the merger of inner binary stars in hierarchical triple systems has been proposed \citep{perets2009triple, naoz2014mergers}. While these mechanisms have garnered strong support from the various evidences, it is worth noting that there are certain BSS that cannot be accounted for by these models \citep{cannon2014blue}. 

However, the phenomenon of BSS being brighter and bluer than MSTO is possible only when the accreted mass is sufficient to make the progenitor star brighter than the MSTO and/or the progenitor is an MSTO star. If the amount of accreted gas is insufficient or the accretor star is too small, the shift in the CMD would not be significant enough for the accretor to surpass the brightness of the MSTO. These objects will share similar characteristics with BSS but cannot be classified as such since they do not exhibit a bluer and brighter profile than the MSTO. As a result, these stars have been referred to as BLs \citep{leiner2019blue}. Although the detection of BLs poses a challenge as they resemble typical main sequence stars, several techniques have been developed to identify them. These techniques include observing higher-than-average rotation. Since BLs often exhibit elevated rotation rates, this can serve as an indicator of recent MT events. Secondly, the identification of a companion star proves the mass donation. Finally, the presence of chemical peculiarities serves as a method to identify the BLs since they display peculiar chemical compositions.

YSS, which are another intriguing type of stellar population, can be observed in optical CMDs positioned on the blue side of the red giant branch and above the sub-giant branch. A study conducted by \cite{mathieu1986spatial} focused on YSS in the open cluster M67 and revealed that the majority of these stars were spectroscopic binaries and were identified as members of the cluster based on their proper motion and radial velocities. \cite{landsman1997s1040} discovered a YSS in the same cluster with a white dwarf (WD) companion, indicating a history of MT. Additionally, \cite{leiner2016k2} concluded that a merger or collision event was likely responsible for the YSS, suggesting that these could be evolved BSS. Consequently, similar to BSS, these objects are believed to have formed via various mechanisms. On the other hand, RC stars are cool horizontal branch stars that have experienced a helium flash and are currently undergoing helium fusion in their cores. As a result, they exhibit red colors and are situated near the red giant branch in the CMD \citep{girardi2016red}.

UV wavelengths provide a distinct benefit to discover the hot companions of the above mentioned objects through the detection of excess UV wavelength emissions. For example, \cite{gosnell2015implications} successfully identified WD companions in 7 out of 15 BSS in the open cluster NGC 188 by utilizing FUV data acquired from the Hubble Space Telescope (HST). Similarly, \cite{rao2022characterization} discovered a low-mass (LM) WD as the hot companion of a BSS in open cluster Melotte 66 using data from Ultraviolet Optical Telescope (UVOT).

Apart from playing a significant role in the formation of the above-mentioned exotic stellar populations, binary stars are crucial in the dynamics of stellar systems, spanning from small open star clusters \citep{hut1992binaries} to dwarf spheroidal galaxies \citep{spencer2018binary}. It is believed that a significant fraction of these binaries are primordial in nature, and the presence of equal-mass binary systems within a star cluster significantly influences its dynamical evolution \citep{hut1992binaries, portegies2001star, binney2008galactic, kouwenhoven2008effect}. Binary population synthesis calculations, although simplified, have demonstrated that binary evolution can give rise to a subset of stars that exhibit rapid rotation \citep{de2013rotation}. 

As mentioned above, the majority of the clusters, spanning ages from approximately 700 Myr to 2 Gyr, exhibit a feature termed eMSTO. This phenomenon has been observed in studies by \cite{mackey2007double, milone2009multiple, goudfrooij2011population, goudfrooij2014extended}, among others. \cite{cordoni2018extended} examined 12 Galactic open clusters and found compelling evidence that the eMSTO is not an exclusive characteristic limited to star clusters in the Magellanic Clouds. Instead, it is a common feature observed in Galactic open clusters as well. Initially, the presence of the eMSTO feature was thought to be a result of a significant age spread, spanning hundreds of millions of years \citep{mackey2007double, milone2009multiple, goudfrooij2011population, goudfrooij2014extended}. However, subsequent research has indicated that a stellar evolutionary effect, specifically stellar rotation, is the more likely cause \citep{bastian2009effect,brandt2015rotating,niederhofer2015apparent,cabrera2016escape}. The idea of actual age spreads as the origin of the eMSTO has been less favored due to the observed correlation between the extent of the inferred age spread and the cluster's age \citep{niederhofer2015apparent}. Spectroscopic studies of stars located on the eMSTO in Galactic open clusters have provided evidence supporting the link between stellar rotation and their position within the MSTO \citep{dupree2017ngc, kamann2018cluster, bastian2018extended, marino2018discovery}. 
The observational evidence that rapidly rotating stars mostly lie on the red side of the eMSTO, whereas blue eMSTO stars are typically slow rotators \citep{bastian2018extended, marino2018discovery} is in contrast with the expectations if rotation is only responsible for the eMSTO \citep{d2015extended, milone2017multiple}. The recent work by \cite{d2023role} has provided a novel explanation for the eMSTO feature. Their analysis of the eMSTO in a Large Magellanic Cloud cluster NGC 1783 shows that it can be explained by absorption due to the circumstellar dust present around the stars. The effect of the circumstellar dust is shown to be modulated by the inclination angle of the rotation axis, such that the stars more affected by the dust are redder.

NGC 6940 is a well studied intermediate age ($\sim$1 Gyr) open cluster \citep{bocek2016chemical}, located at a distance of $\sim$ 770 pc \citep{kharchenko2005astrophysical}. 
The initial investigation of cluster membership of this cluster through relative proper motion was carried out by \cite{vasilevskis1957relative}. 
\cite{twarog1997some} included NGC 6940 in their study on photometric and spectroscopic data of 76 open clusters and derived a metallicity of \big[Fe/H\big] = 0.01 $\pm$ 0.06. \cite{friel2002metallicities} calculated the metallicity using the spectroscopic data from \cite{thogersen1993metallicities} and obtained \big[Fe/H\big] = $-$0.12 $\pm$ 0.10. \cite{mermilliod2008red} observed 26 potential red giants and identified five non-RV members. They updated their previous RV measurements and obtained a cluster velocity of RV = 7.89 $\pm$ 0.14 km/s. In a recent high-resolution spectroscopic study of 31 open clusters, \cite{blanco2015testing} included one red giant member from NGC 6940. Their analysis yielded \big[Fe/H\big] = +0.09 $\pm$ 0.07. Collectively, these studies indicate that NGC 6940 exhibits an overall metallicity close to that of the solar.

Despite extensive photometric and spectroscopic studies conducted on NGC 6940, the exotic stellar populations mentioned earlier have not been thoroughly examined in this cluster. Moreover, the presence of eMSTO in this cluster makes it an interesting candidate for investigation. This cluster has been observed with the Ultraviolet Imaging Telescope (UVIT) on board the \textit{AstroSat}, which has the ability to detect exotic objects and their hot companions when combined with other multi-wavelength data. Several studies have explored this aspect in multiple open clusters. For instance, \cite{subramaniam2016hot} identified a post-AGB/horizontal-branch companion of a BSS in the open cluster NGC 188 using UVIT. \cite{sindhu2019uvit}, \cite{jadhav2019uvit}, and \cite{pandey2021uocs} investigated the open cluster M67 and discovered extremely-low mass (ELM; $<$ 0.2 M$_{\odot}$) WDs as companions of BSS. Similarly, \cite{vaidya2022uocs} found ELM WDs as hot companions to BSS in the intermediate-age open cluster NGC 7789. In the case of NGC 2506, \cite{panthi2022uocs} detected two ELM WDs and one LM WD (with masses ranging between 0.2 M$_{\odot}$ and 0.4 M$_{\odot}$) as the hot companions of BSS, and normal-mass WD (with masses ranging between 0.4 M$_{\odot}$ and 0.6 M$_{\odot}$) and high-mass ($>$ 0.6 M$_{\odot}$) WDs as hot companions of YSS and RC stars. \cite{rani2023uocs} reported A-type sub-dwarf as the hot companion of YSS in the open cluster NGC 2818. Recently, \cite{panthi2023uocs} identified ELM, LM, normal-mass, and high mass WD companions of the BSS in the open cluster NGC 7142.

In this study, we present the first-ever analysis of BSS, BLs, YSS, and RC stars of NGC 6940 using the UVIT data, with the objective of shedding light on their formation mechanisms. Additionally, we investigate the influence of equal-mass binary stellar populations on the dynamical evolution of the cluster. Furthermore, we examine the contribution of age spread and rotation on the observed eMSTO phenomenon. This paper is organized as follows: \S \ref{Section 2} provides details on the observations and data reduction procedures. The methodology employed in this study is described in \S \ref{Section 3}. In \S \ref{Section 4}, we present the results obtained, followed by a comprehensive discussion in \S \ref{Section 5}. Finally, \S \ref{Section 6} provides the summary and conclusions of the work.

\section{Observations and data reduction} \label{Section 2}

NGC 6940 was observed using UVIT on 13$^{th}$ June 2018 in a single FUV filter, F169M. These observations were performed under the proposal ID A04-075 with an exposure time of $\sim$2000 s. UVIT is one of the payloads onboard \textit{AstroSat} (the first multi-wavelength Indian Space Observatory) and serves as the ultraviolet and optical observation instrument \citep{singh2014astrosat}. It comprises of two Ritchey-Chretien telescopes with a primary mirror of diameter $\sim$38 cm. The instrument offers a circular field of view with a diameter of about 28$\arcmin$. The spatial resolution is $\sim$ 1.5$\arcsec$ for both the FUV and NUV channels. Each channel is equipped with a selection of filters that offer different wavelength passbands \citep{tandon2017orbit, tandon2020additional}. For additional details regarding the effective area curves, calibration results, and instrumentation, readers are referred to \cite{kumar2012ultraviolet} and \cite{tandon2017orbit, tandon2020additional}. To process the Level 1 data and generate scientific images, we utilized a customized software package called CCDLAB \citep{postma2017ccdlab}. This software performs various corrections, including geometric distortion, centroiding bias, flat-field illumination, and spacecraft drift. The resulting images after drift correction are aligned and combined to produce science-ready images. In order to perform astrometry, we applied the coordinate matching algorithm using \textit{Gaia} DR2, which is incorporated into the upgraded version of CCDLAB \citep{postma2020algorithm}.

\section{Methodology} \label{Section 3}

\subsection{Determination of cluster members}

Ensuring secure membership in open clusters is crucial, as contamination from field stars poses a significant challenge. To address this, we employ the ML-MOC algorithm developed by \cite{agarwal2021ml} using \textit{Gaia} DR3 data. This machine-learning-based approach allows us to determine cluster members without requiring prior knowledge about the cluster. The algorithm combines the k-Nearest Neighbour (kNN, \citealt{cover1967nearest}) and Gaussian mixture model (GMM, \citealt{mclachlan2003clustering}) techniques.

\begin{figure*}
\includegraphics[width=0.35\textwidth]{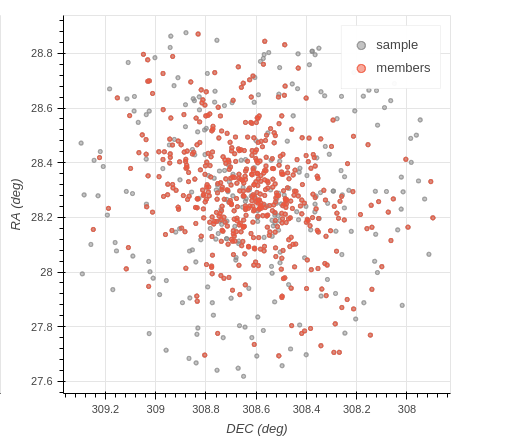}
\includegraphics[width=0.3\textwidth]{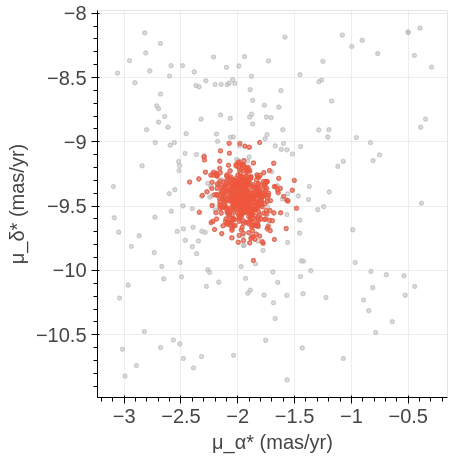}
\includegraphics[width=0.3\textwidth]{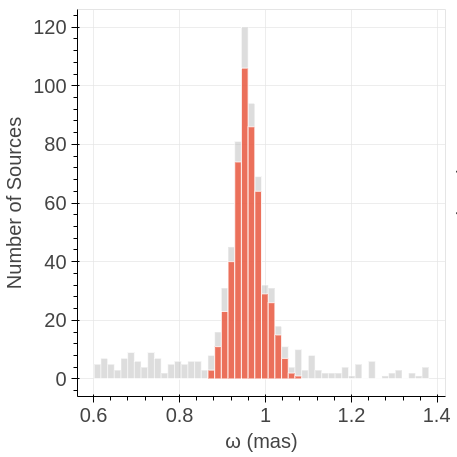}
\caption{The spatial distribution, proper motion, and parallax distributions of the members and sample sources determined using the ML-MOC algorithm.}
\label{Fig.1}
\end{figure*}

To determine cluster members using the ML-MOC algorithm, we followed these steps. Initially, we download sources within 40$\arcmin$ radius centered on the cluster, with five astrometric parameters (positions, proper motions, and parallax), appropriate measurements in the Gaia photometric passbands G, G$_{BP}$, and G$_{RP}$, non-negative parallaxes, and having an error in G-magnitude less than 0.005. These sources were referred to as \textit{All sources}. Next, we estimate the mean proper motions and mean parallaxes of the cluster by utilizing the kNN algorithm. This allows us to determine the ranges of proper motions and parallaxes that encompass all likely cluster members. These sources were labeled as \textit{Sample sources}. Subsequently, we apply the GMM, an unsupervised clustering algorithm, to the \textit{Sample sources}. By fitting two Gaussian distributions in the proper motion and parallax space, we aim to separate the probable cluster members from the field stars. The GMM also assigned membership probabilities to the sources. Our initial members are high-probability members with membership probabilities greater than 0.6. Later, we include \textit{Sample sources} with membership probabilities between 0.2 and 0.6 in the list of member sources. We identified 492 sources as cluster members. The proper motion, spatial, and parallax distributions of the \textit{Sample sources}, as well as the identified cluster members, are illustrated in Figure \ref{Fig.1}. 

\subsection{UV Photometry}

In order to get the FUV magnitude of the stars, we performed Point Spread Function (PSF) photometry using DAOPHOT package of Image Reduction and Analysis Facility (IRAF, \citealt{stetson1987daophot}). The process of performing PSF photometry includes the following tasks: First, we utilize the \textit{imexam} task to estimate the background counts and the FWHM of the image. Next, we use the \textit{daofind} task to identify the sources that surpassed a specific threshold, typically 3-10 times the background level. Following the identification of sources, we perform the \textit{phot} task in order to do aperture photometry. Then we use \textit{pstselect} task to choose isolated stars, followed by the \textit{psf} task to generate a PSF model. Finally, we execute the \textit{allstar} task to estimate PSF magnitudes for all stars simultaneously, employing multiple iterations of PSF fitting to groups of stars. Additionally, we implemented PSF corrections to mitigate any systematic differences between PSF photometry magnitudes and aperture photometry magnitudes. To determine the aperture correction, we use the curve-of-growth technique. Furthermore, it is important to consider that the FUV detector of UVIT operates in photon counting mode, which means it may encounter counts exceeding one photon per frame. To address this issue, we applied saturation corrections \citep{tandon2020additional}. This allowed us to eliminate the effects of saturation and ensure accurate measurements. 

\begin{figure}
\includegraphics[width=0.5\textwidth]{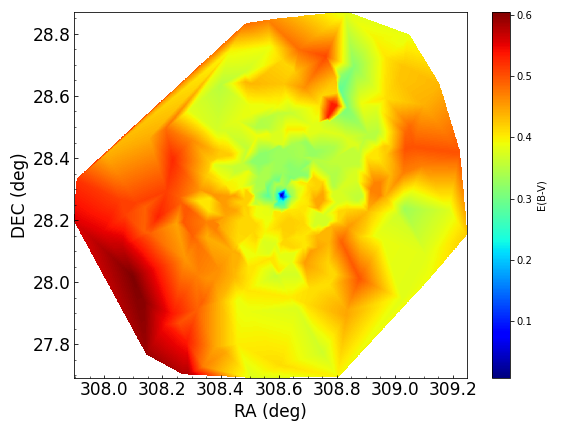}
\includegraphics[width= 0.45\textwidth]{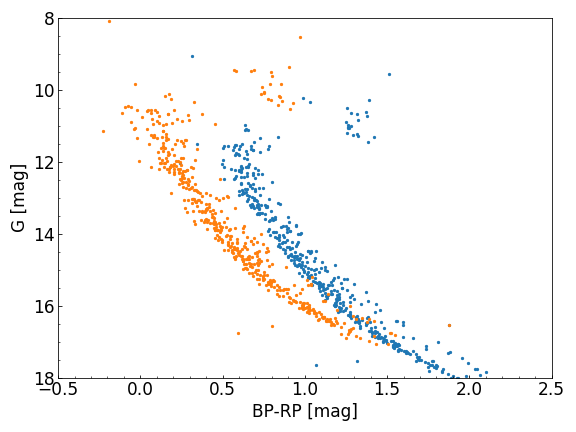}
\caption{The upper panel shows the reddening map of the cluster members, generated by 3D dust maps \citep{green20193d}. The lower panel shows the observed CMD of Gaia DR3 members as blue dots and the differential reddening corrected CMD of Gaia DR3 members as orange dots.}
\label{Fig.2}
\end{figure}

\begin{figure}
\includegraphics[width=\columnwidth]{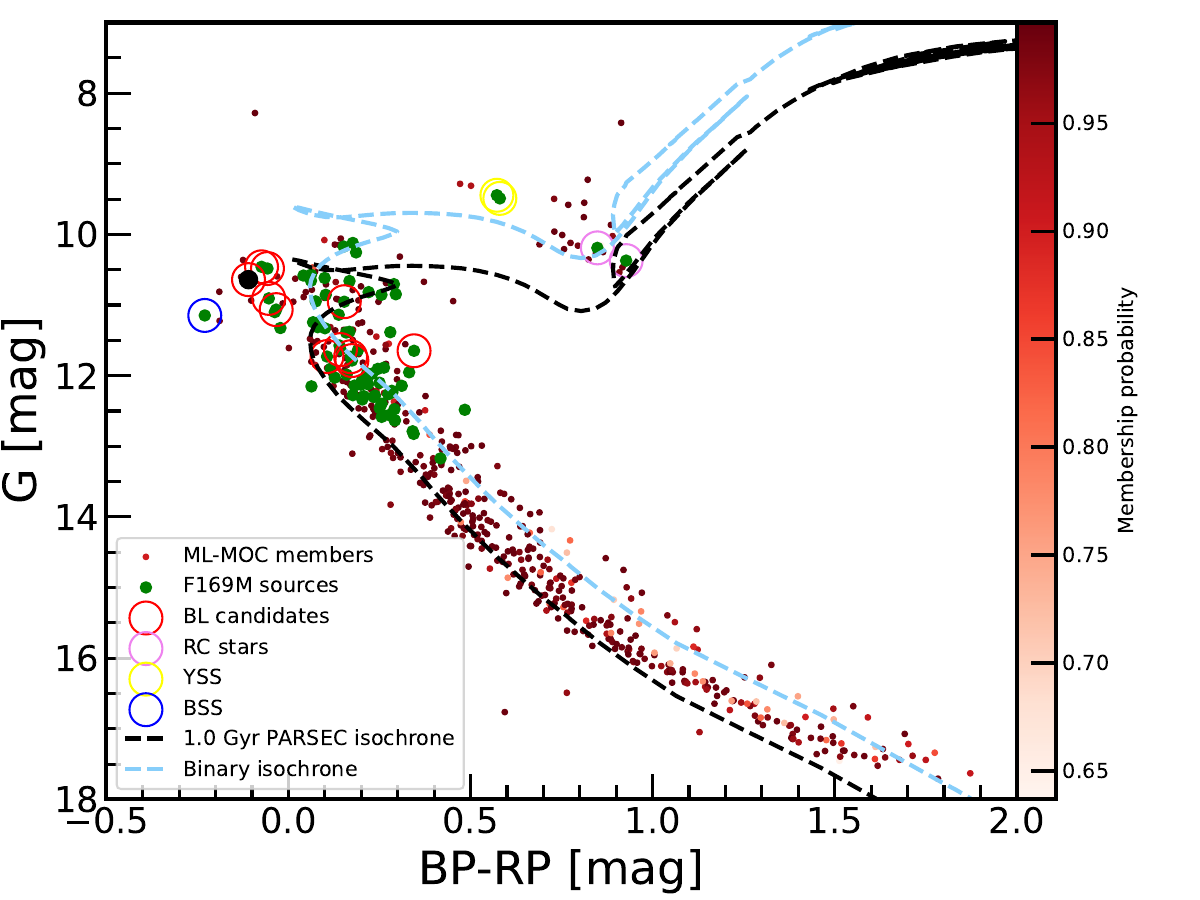}
\caption{The differential reddening corrected optical CMD of NGC 6940 showing members determined using ML-MOC. The color bar represents the membership probability of each source. All objects observed in the UVIT/F169M filter are shown as green dots. The following symbols are used to mark different populations: BSS as blue open circle, BL candidates as red open circles, YSS as yellow open circles, and RC stars as pink open circles. The BL shown as a black dot appears as a BSS in the observed CMD.} A PARSEC isochrone \citep{bressan2012parsec} of age 1.0 Gyr, \big[Fe/H\big] = 0 \citep{blanco2015testing}, after applying the extinction correction of A$_{\text{G}}$ = 0.66 and reddening of E(B$_{\text{P}}$ $-$ R$_{\text{P}}$) = 0.214 \citep{loktin2001catalogue} is shown as a black curve, and an equal-mass binary sequence isochrone derived by adding 0.75 mag to the single star isochrone is shown as a blue dashed curve.
\label{Fig.3}
\end{figure}

\section{Results} \label{Section 4}

\subsection{The color-magnitude Diagram}

In order to quantify and correct the differential reddening that may be present in this cluster, we used 3D dust maps \citep{green20193d}. The upper panel of Figure \ref{Fig.2} shows the reddening map of the cluster generated using the 3D dust maps. It can be seen that in most parts, the reddening, E(B-V), varies from $\sim$0.3 to $\sim$0.6, with the maximum reddening effect observed in the south-west. The BSS, YSS, and BLs studied in this work are located in the central regions, suffering average reddening, E(B-V) $\sim$0.4 $-$ 0.5. The lower panel of  Figure \ref{Fig.2} shows the observed CMD and the differential reddening corrected CMD. It is noteworthy that even on applying the differential reddening correction, the broadening of the upper main sequence region is not reduced. It implies that this broadening is not due to the differential reddening, but the cluster has an eMSTO.

In Figure \ref{Fig.3}, we have shown the differential reddening corrected optical CMD, with the membership determined using the ML-MOC algorithm. We identified one BSS candidate observed in the UVIT filter. Sources near the MSTO with high rotational velocities (vsin\textit{i} $\sim$50 km/s or greater) in \textit{Gaia} DR3 and a counterpart in F169M are considered BL candidates. We identified 11 such BL candidates. One of these BLs (marked as a black dot) could be considered a BSS candidate based on the observed CMD. However, on applying the differential reddening correction to the cluster, this source gets shifted towards the main sequence. Hence, we consider it to be a BL candidate. Sources situated above the SGB region and bluer compared to the RGB are classified as YSS candidates. Both these YSS candidates also have an F169M counterpart. We found two RC stars that have counterparts in the F169M filter. These BSS, BLs, YSS, and RC candidates are marked as open circles in Figure \ref{Fig.2}. In addition, all members with a counterpart in the F169M filter are marked as green dots.

\subsection{Radial density profile}

\begin{figure}
\centering
\includegraphics[width=\columnwidth]{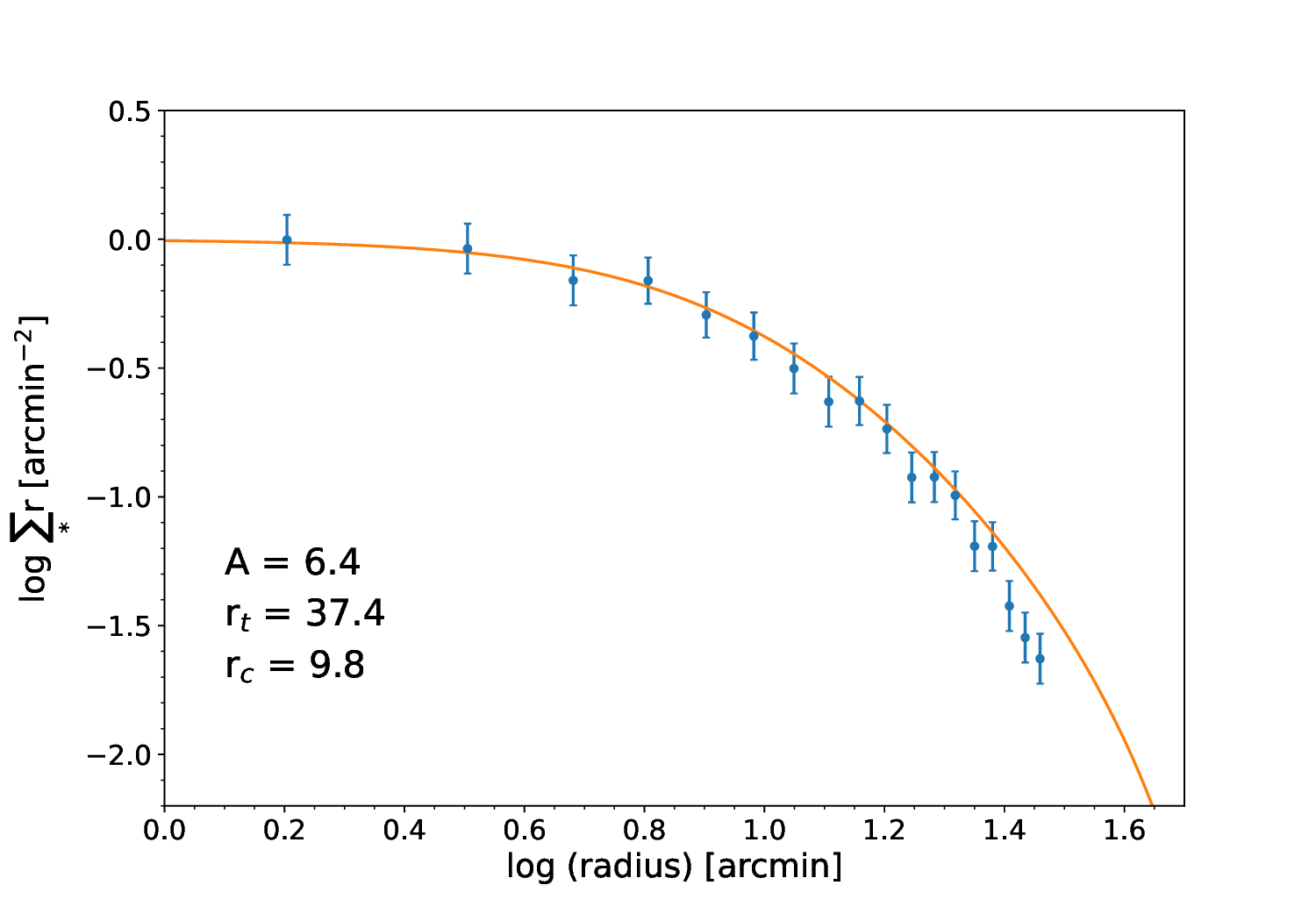}
\caption{The radial density profile of the cluster members, where the error bars represent 1 $\sigma$
Poisson errors. The resulting parameters from King's function fit, including core radii, tidal radii, and the normalization factor A, are indicated at the bottom left corner.}
\label{Fig.4}
\end{figure}

We fit the radial density profile of the cluster members using King's function \citep{king1962structure} as shown in Figure \ref{Fig.4}. The fitting process involves dividing the range of cluster radii into $\sim$ 25 equal-radius bins. For each bin, we plot the logarithm of the number density against the radius. Notably, the King's function exhibited a good fit to the cluster data. We note that the core radius r$_{c}$ is $\sim$9.8 $\arcmin$, whereas the tidal radius r$_{t}$ is $\sim$37.4 $\arcmin$. The structural parameters we derive are in agreement with those reported for the cluster in \cite{cordoni2023photometric}.

\subsection{Mass segregation}

In a star cluster, different stellar interactions occur, which lead to various dynamical processes including mass segregation \citep{meylan1997internal}. To know the dynamical status of this cluster, we analyze the spatial distributions of various populations such as low-mass single stars, massive single stars, and equal-mass binaries. A commonly used approach to study equal mass binaries in the CMD involves dividing the CMD into distinct regions for single stars and binary systems and then matching the number of cluster members within each respective region. This methodology has been applied in various studies concerning star clusters, such as those conducted by \cite{sollima2010fraction, milone2012acs, li2013binary}, and the reference therein. We segregate the main-sequence equal-mass binaries by visual inspection of the CMD as shown in Fig \ref{Fig.3}. We classify the single star populations with magnitudes in the G-band $<$ 14 as massive stars and magnitudes in the G-band $>$ 14 as low-mass stars. 

\begin{figure}
\includegraphics[width=\columnwidth]{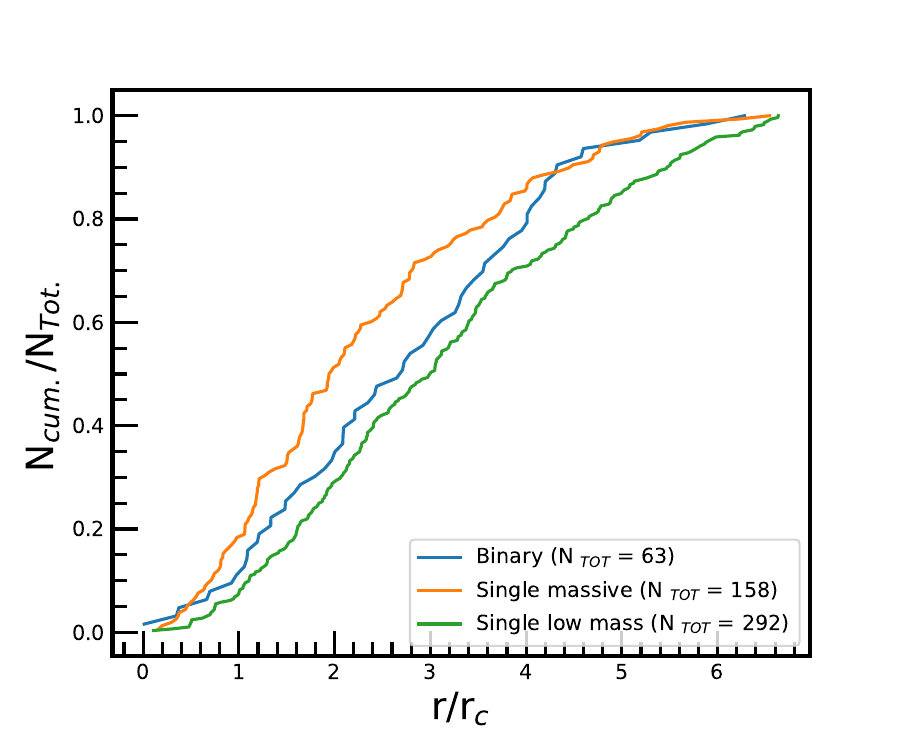} 
\caption{The cumulative radial distributions of equal-mass binary, massive single, and low-mass single populations. The normalized number of the three populations is plotted on the y-axis, and the radial distance in the units of core radius, r$_{c}$, is plotted on the x-axis.}
\label{Fig.5}
\end{figure}

In Figure \ref{Fig.5}, we show the cumulative radial distributions of the low-mass single, massive single, and equal-mass binary main sequence populations within the cluster. The y-axis represents the normalized numbers of populations, while the x-axis represents the radial distance in units of r$_{c}$. The plot reveals significant segregation of massive single stars compared to low-mass single stars, indicating the dynamical evolution of the cluster. Being among the massive objects within the clusters, equal-mass binary stars are anticipated to be strongly influenced by the phenomenon of dynamical friction, which leads to their segregation towards the center of the cluster. The plot indeed shows the equal-mass binaries to be more concentrated as compared to the lower-mass single stars.

\subsection{The extended main sequence turn-off}

\begin{figure}
\includegraphics[width=\columnwidth]{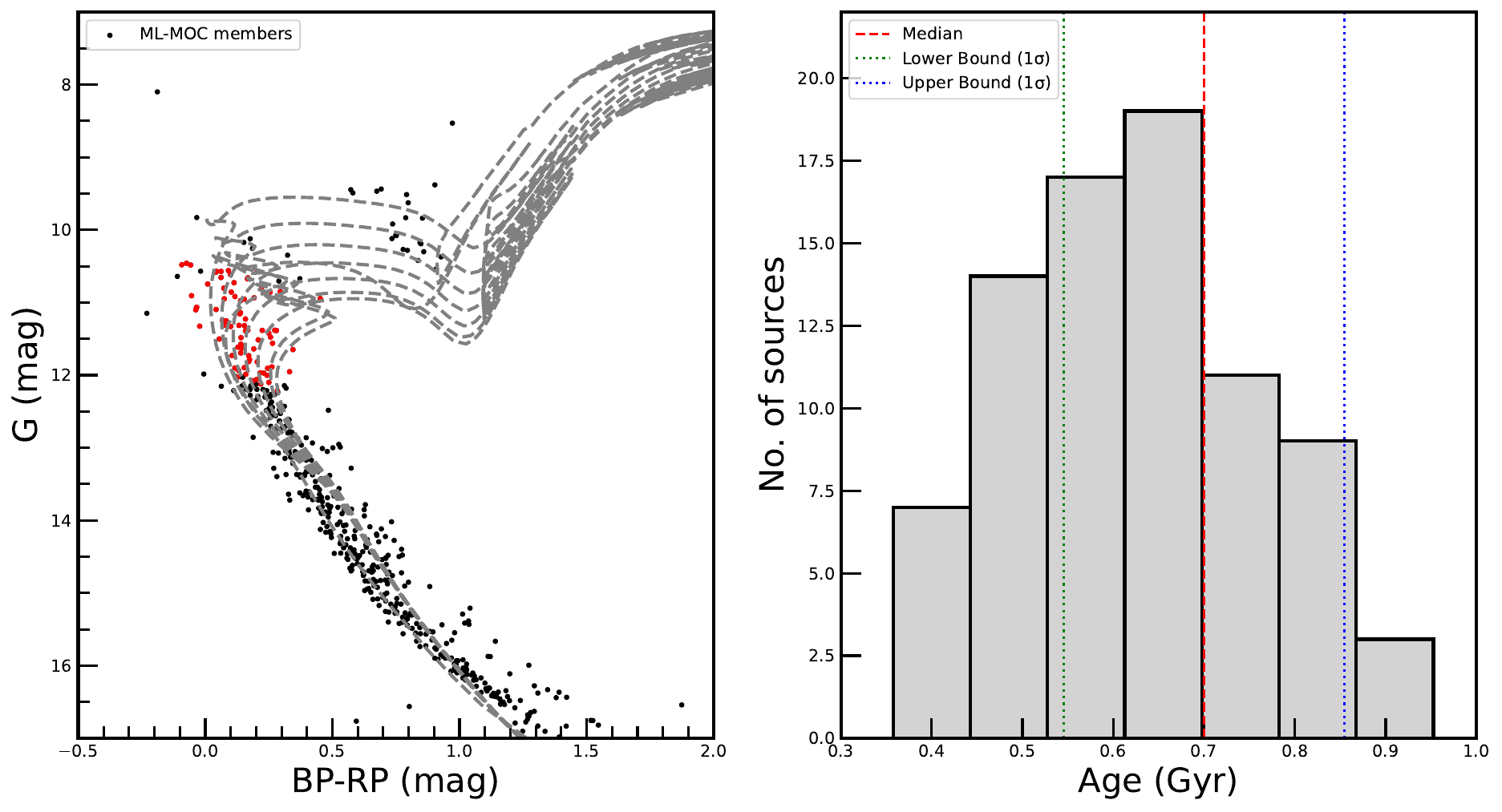}
\caption{The left panel shows grid of isochrones ranging in age from 0.4 to 1 Gyr, with steps of 0.1 Gyr over-imposed on the differential reddening corrected CMD. The right panel shows the histogram of age distribution of the eMSTO stars (red dots) in the left-panel CMD. The median age of these stars is marked with a vertical continuous line, while the two dashed lines have distances of 1 $\sigma$ from the median value.}
\label{Fig.6}
\end{figure}

To comprehend the underlying cause of the observed eMSTO in this cluster, we undertake an investigation following  \cite{cordoni2018extended}, where we mark the sources considered for investigation of age differences as red dots in Figure \ref{Fig.6}. Only these sources were utilized to infer the age distribution due to eMSTO. Subsequently, a grid of isochrones, with the same metallicity, was overlaid onto the differential reddening corrected CMD. These isochrones ranged in age from 0.4 to 1 Gyr, with steps of 0.1 Gyr. Next, we associate each star with the age of the closest isochrone, allowing us to obtain the age distribution displayed in the right panel of the figure. Finally, we calculate the median age to be $\sim$0.7 Gyr, and the age spread considering 1 $\sigma$ of the distribution of these absolute values to be $\sim$0.15 Gyr. In literature, clusters of comparable age exhibiting eMSTO show an age spread of 70 Myr-250 Myr \citep{cordoni2018extended}. The age spread we observe in this cluster is not negligible. Therefore, we cannot rule out the possibility of age spread as one of the factors contributing to the observed eMSTO.

\begin{figure}
\includegraphics[width=\columnwidth]{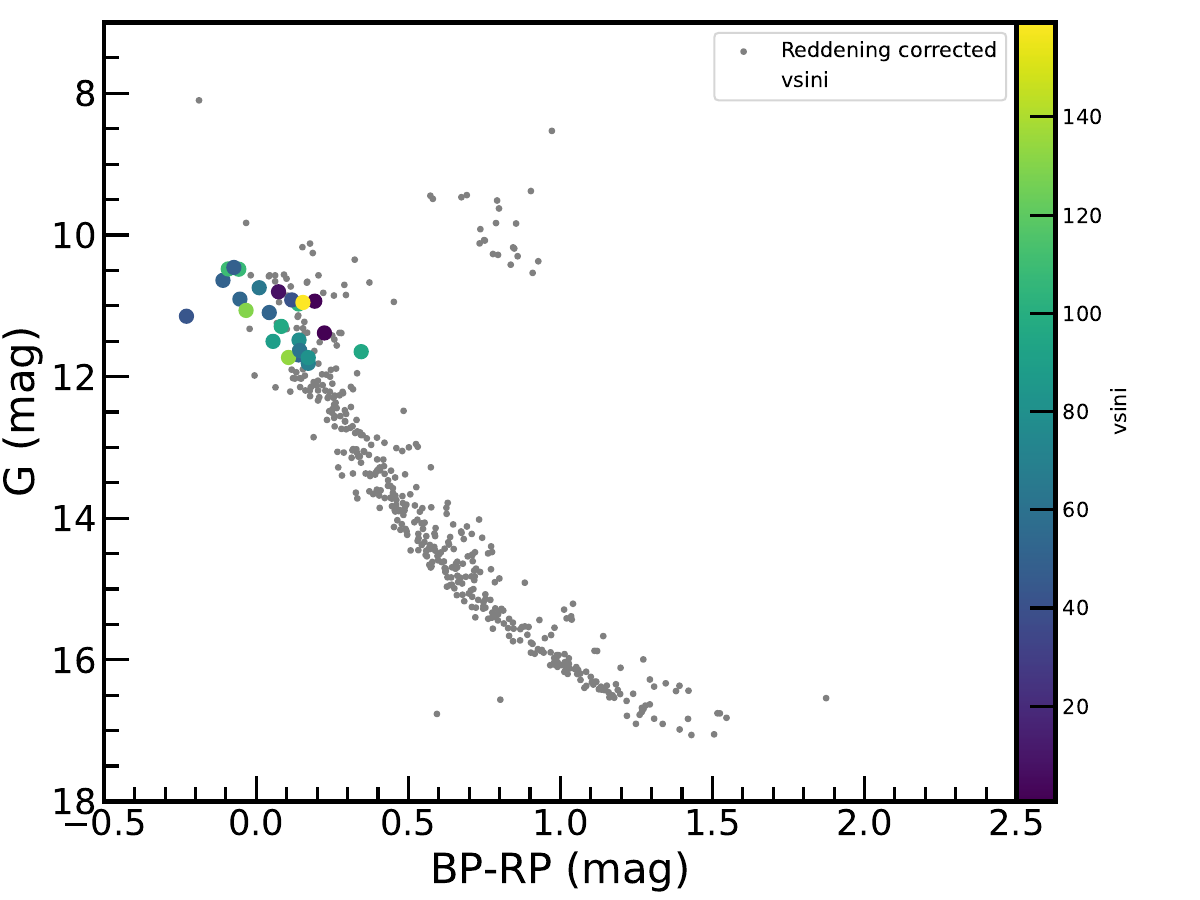}
\caption{Differential reddening corrected optical CMD with rotational velocities from \textit{Gaia} DR3 represented in the color-bar.}
\label{Fig.7}
\end{figure}

On the other hand, it is also possible that the apparent wide age spread inferred from the eMSTO is an artifact due to stellar rotation \citep{cordoni2018extended,cordoni2022ngc1818}. To investigate whether the phenomenon of eMSTO is linked to rotation, we checked the available literature. We found that the values of vsin\textit{i} are available in \textit{Gaia} DR3 for a small fraction of sources along MSTO  \citep{gaia2022vizier}, and they vary from $\sim$10 km/s to 150 km/s, as depicted in the differential reddening corrected CMD in Figure \ref{Fig.7}. As anticipated, stars with higher rotational velocities were found to occupy brighter magnitudes, while those with lower rotational velocities were distributed at fainter magnitudes \citep{cordoni2018extended}. As mentioned above, if rotation is indeed the underlying cause of the eMSTO phenomenon, we would expect rapidly rotating stars to predominantly appear on the redder side of the eMSTO, while slower rotating stars would be more likely to reside on the bluer side \citep{bastian2018extended}. About half of the high rotational speed eMSTO sources (4/7) are, in fact, appearing on the bluer side in this case. We have vsin\textit{i} information only for $\sim$ 12$\%$ of the eMSTO sources, therefore it is not possible to confirm any trend. As recently reported by \cite{d2023role} the eMSTO can be accounted for by dust absorption. The suggested model signifies a major departure in interpreting the eMSTO. The positioning of stars within the turn off correlates with the impact of surrounding dust on their inherent colors. Consequently, stars predominantly influenced by dust will be redder. While the modeling conducted by \cite{d2023role} to investigate the influence of dust on the eMSTO is beyond the scope of this work, we do not dismiss the potential contribution of dust to the eMSTO feature in NGC 6940.

\subsection{Spectral energy distributions}

Stars emit electromagnetic radiation across a broad spectrum of frequencies or wavelengths. The Spectral Energy Distribution (SED) characterizes the distribution of this radiation across different wavelengths. By analyzing the SEDs of stars, we can estimate their fundamental parameters. Moreover, the presence of excess in UV fluxes can provide insights into the potential existence of hot companions, thus offering constraints on their formation mechanisms. We found that two BL candidates have a nearby source within 3\arcsec, and hence we did not construct their SEDs. In order to generate the SEDs of the remaining sources, we utilized the virtual observatory SED analyzer (VOSA, \citealt{bayo2008vosa}) and employed the following steps.

1) We obtain photometric flux measurements for sources using VOSA in the near-ultraviolet (NUV) from GALEX \citep{martin2005galaxy}, optical range from \textit{Gaia} DR3 \citep{babusiaux2022gaia} and PAN-STARRS \citep{chambers2017vizier}, near-infrared from Two Micron All Sky Survey (2MASS, \citealt{cohen2003spectral}), and far-infrared from the Wide-field Infrared Survey (WISE, \citealt{wright2010wide}). \\

2) The photometric fluxes are corrected for extinction by VOSA using the extinction law derived by \cite{fitzpatrick1999correcting} and \cite{indebetouw2005wavelength}. The extinction value A$_{v}$ = 1.55, i.e, E(B-V) = 0.5, obtained from \cite{loktin2001catalogue}, is used for the correction. As mentioned above, the BSS, YSS, and BLs have the reddening $\sim$ 0.4 $-$ 0.5, which is consistent with the average reddening of the cluster, used to fit their SEDs. Table \ref{Table1} presents the extinction-corrected fluxes of all objects in different filters. VOSA obtains synthetic photometry for selected theoretical models by utilizing filter transmission curves and employs a $\chi^{2}$-minimization method to determine the best-fit SED parameters by comparing the synthetic photometry with the extinction-corrected observed fluxes. The reduced $\chi^{2}$ value is calculated using the following formula: \\

\begin{equation} 
   \chi_{r}^{2} =\frac{1}{N-N_{f}}\sum_{i=1}^{N} \frac{(F_{o,i}-M_{d}F_{m,i})^{2}}{\sigma^{2}_{o,i}}
\end{equation} 

In the above equation, N represents the number of photometric data points utilized, while N$_{f}$ denotes the number of free parameters in the model. The observed fluxes of the star are represented by F$_{o,i}$, while F$_{m,i}$ denotes the fluxes predicted by the model. The scaling factor, M$_{d}$, required to achieve the best fit, is determined by multiplying the model by (R/D)$^{2}$, where R represents the radius of the star and D is the distance to the star. Additionally, $\sigma_{o,i}$ signifies the error associated with the observed flux. \\

3) For fitting the SEDs to the objects, we use the Kurucz stellar models \citep{castelli1997notes}. In our analysis, we set the range of effective temperature (T$_{eff}$) from 3500 to 50000 K and set the range of log g from 3 to 5. The \big[Fe/H\big] value was set to zero, which is nearest to the reported value (\big[Fe/H\big] = +0.09) by \cite{blanco2015testing}. \\

\begin{figure*} 
\includegraphics[width=0.45\textwidth]{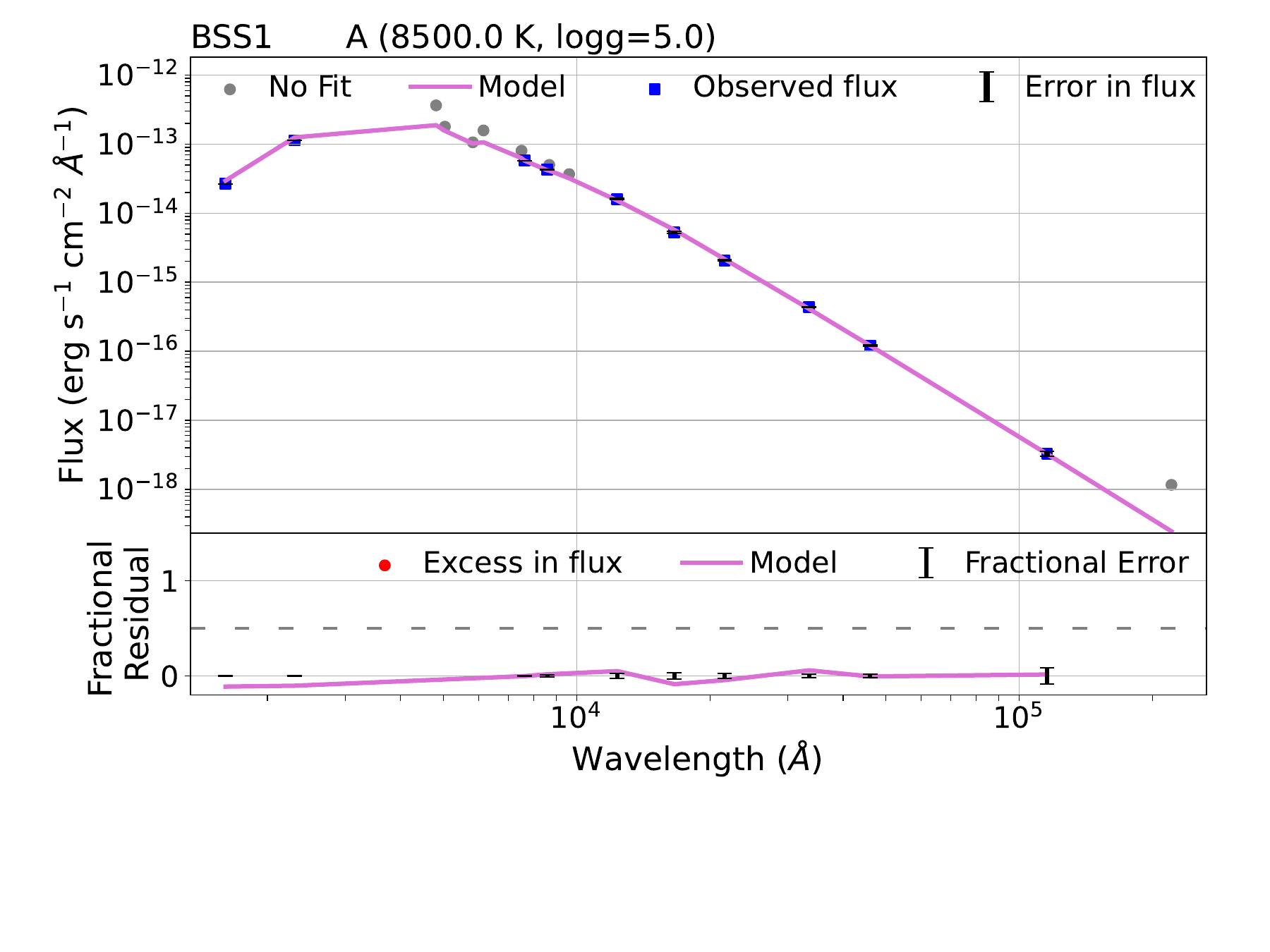}
\includegraphics[width=0.45\textwidth]{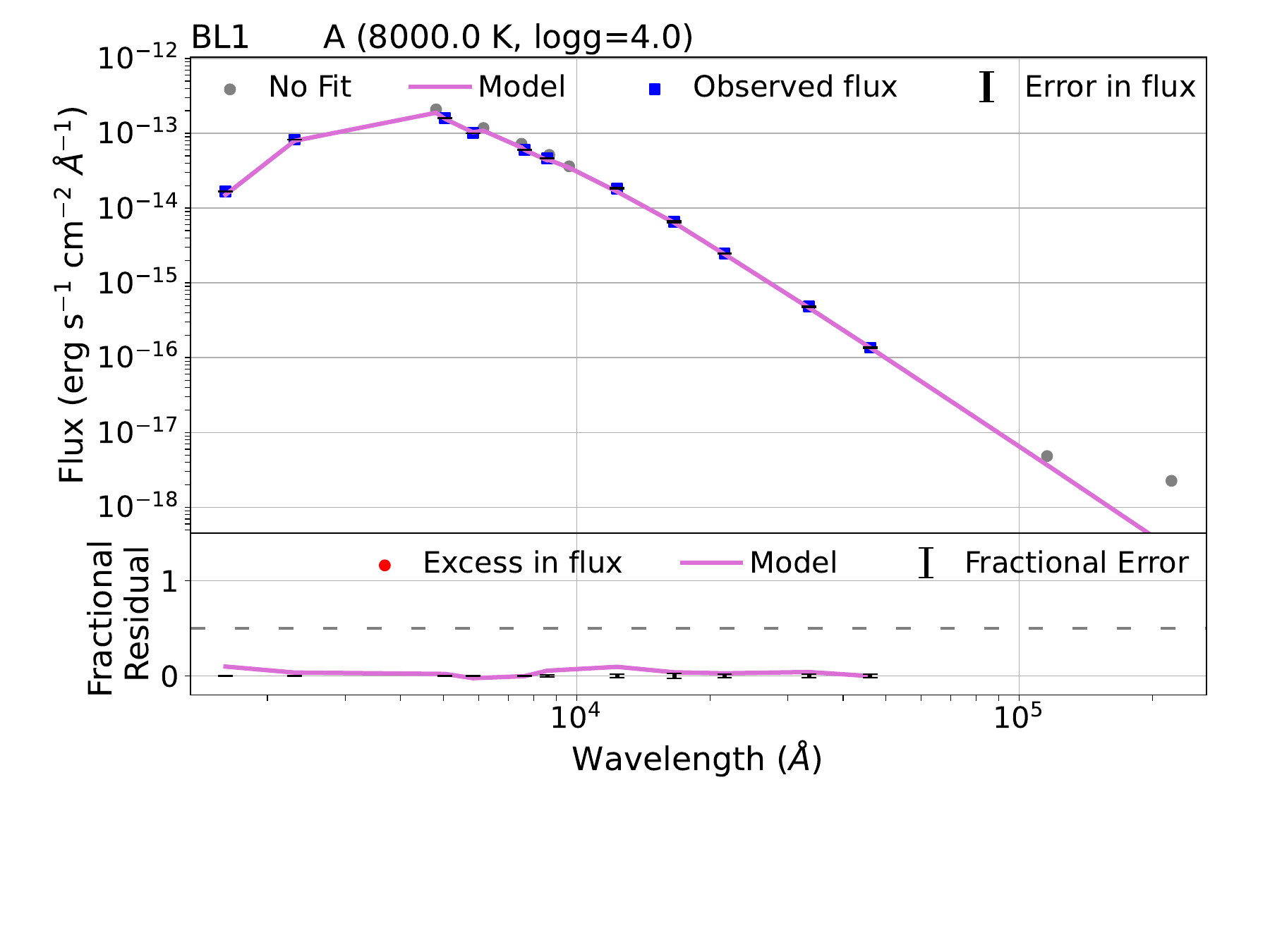} 
\includegraphics[width=0.45\textwidth]{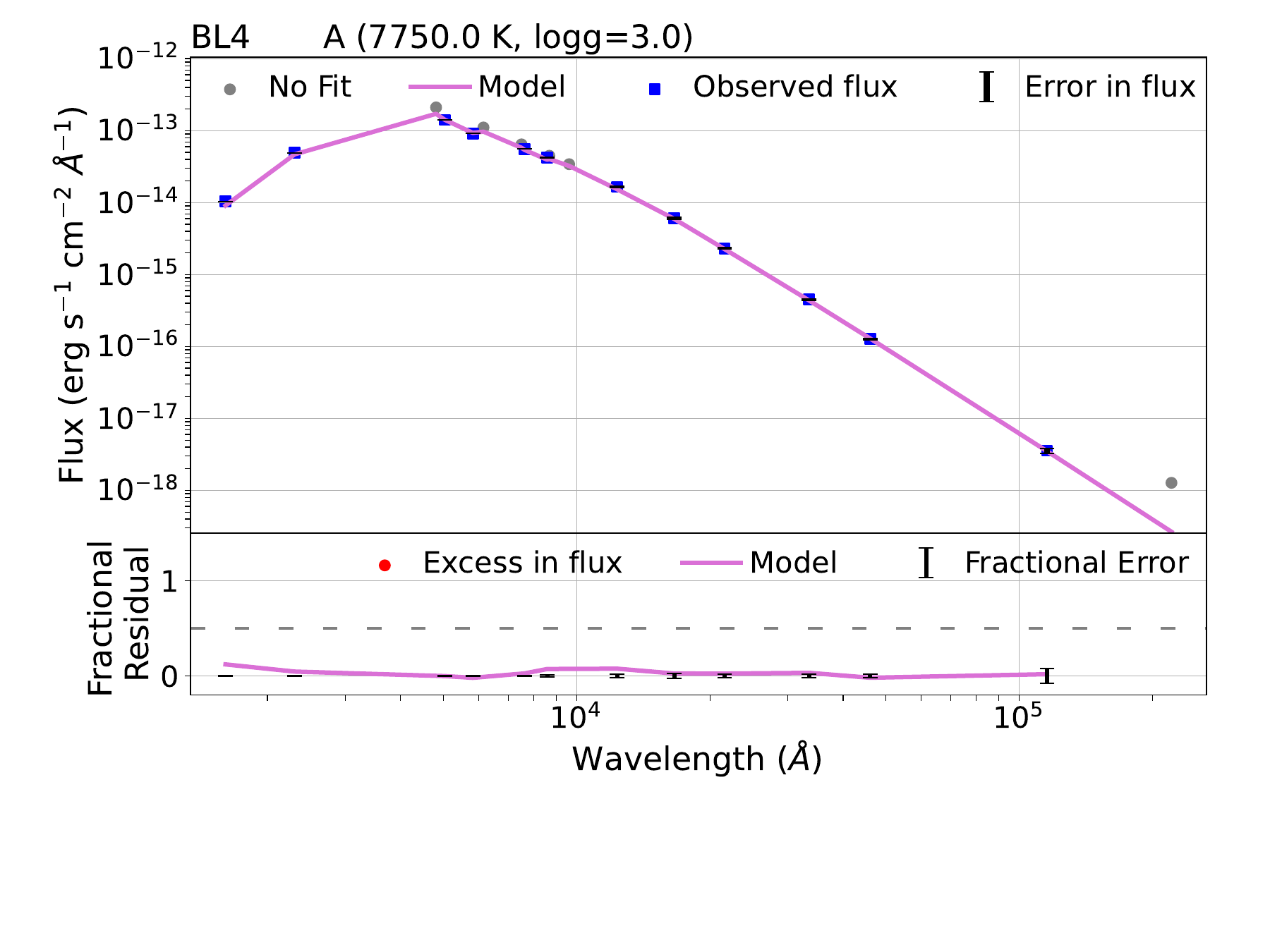} 
\includegraphics[width=0.45\textwidth]{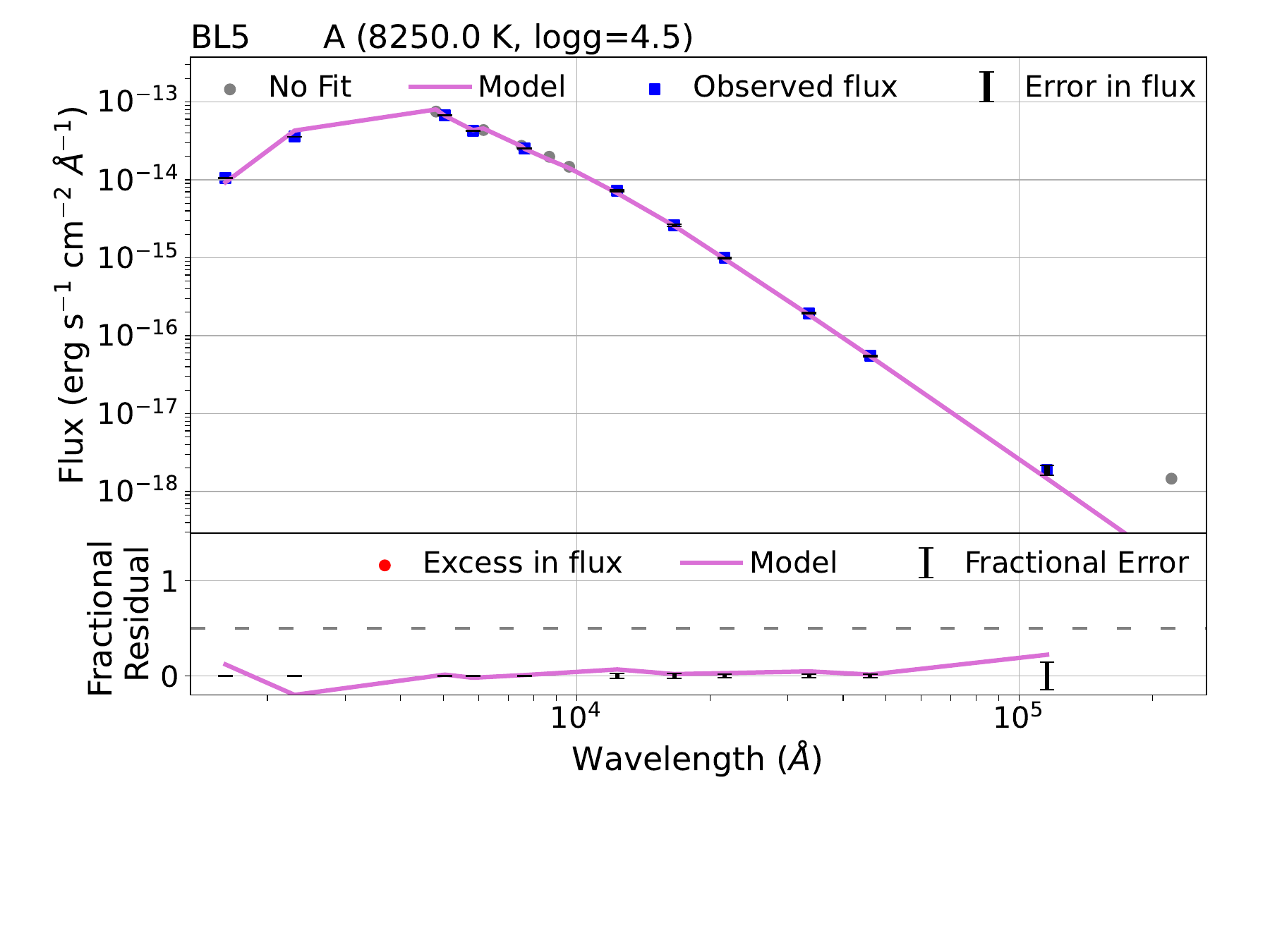}
\includegraphics[width=0.45\textwidth]{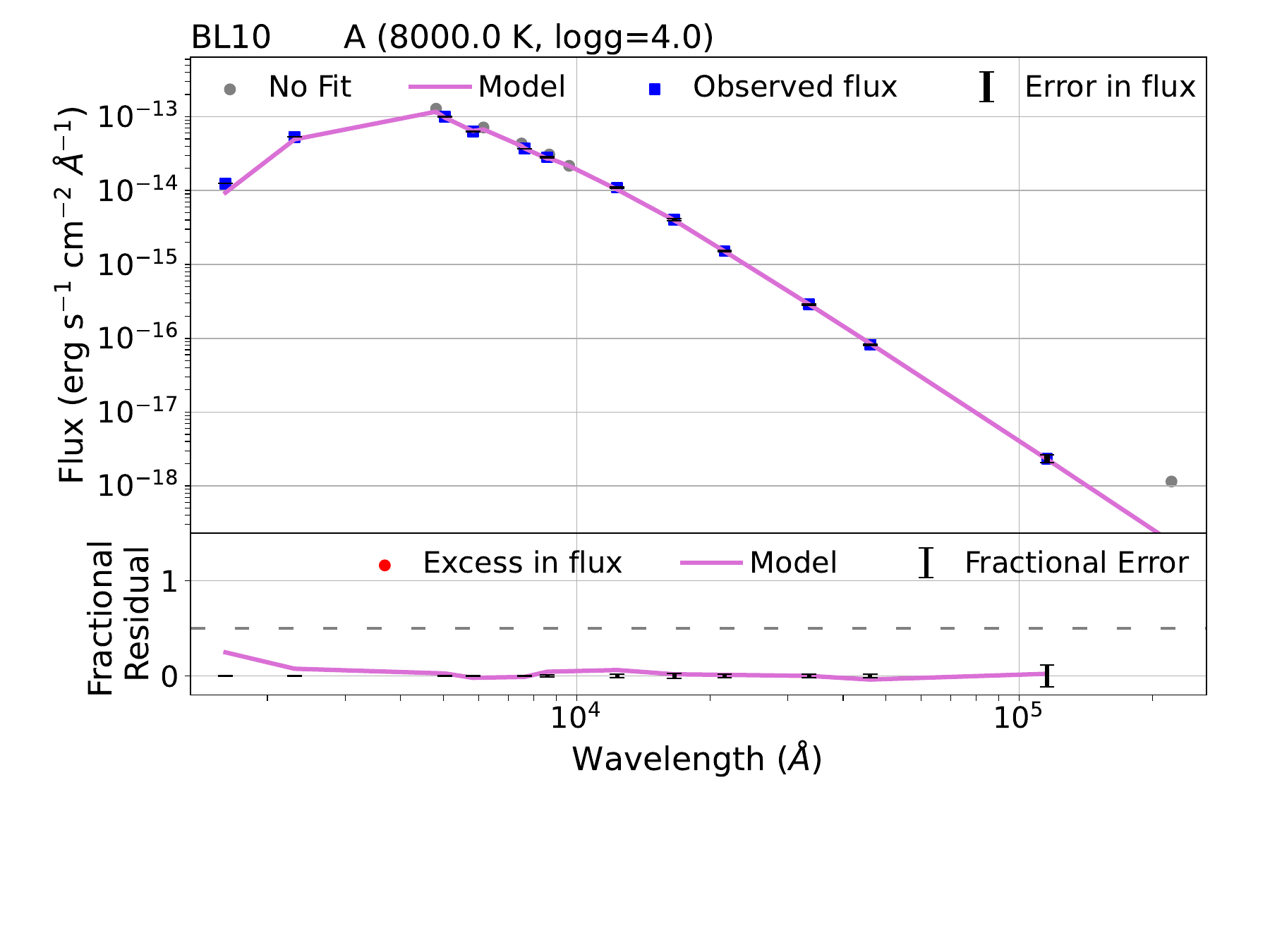} 
\includegraphics[width=0.45\textwidth]{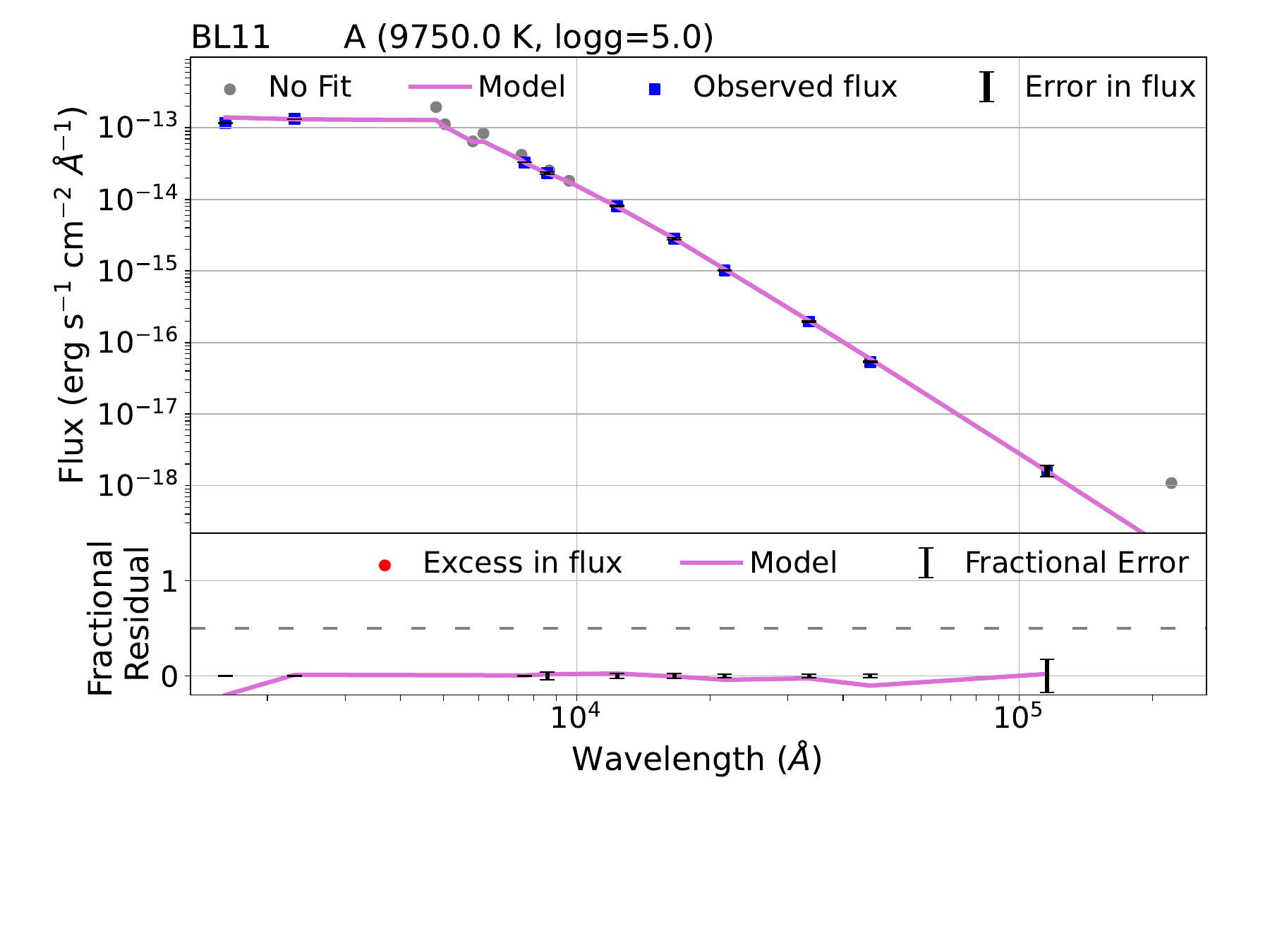}
\caption{Sources fitted with the single component SEDs. The top panel shows blue data points as extinction corrected observed fluxes (labeled as \lq \lq Observed flux''), black error bars representing the errors in observed fluxes (labeled as \lq \lq Error in flux''). The model fit is depicted as a purple curve (labeled as \lq \lq Model''), with grey data points denoting data points that were not included in the fits (labeled as \lq \lq No fit''). The bottom panel shows the residual between extinction-corrected observed fluxes and the model fluxes across the filters from UV to IR wavelengths.}
\label{Fig.8}
\end{figure*}

4) Initially, we exclude the UV data points from the SEDs and check if fluxes fit with the observed fluxes in the optical and IR regimes. We conduct a thorough examination of each UV filter to identify any significant deviations from the expected behavior in relation to the UV and/or IR data points. It is worth mentioning that the BSS and five BLs exhibit a UV flux excess less than 50$\%$. As a result, we fit the single-component SEDs to them. Figure \ref{Fig.8} shows the SEDs of objects fitted with the single-component fit. \\

\begin{figure*} 
\includegraphics[width=0.45\textwidth]{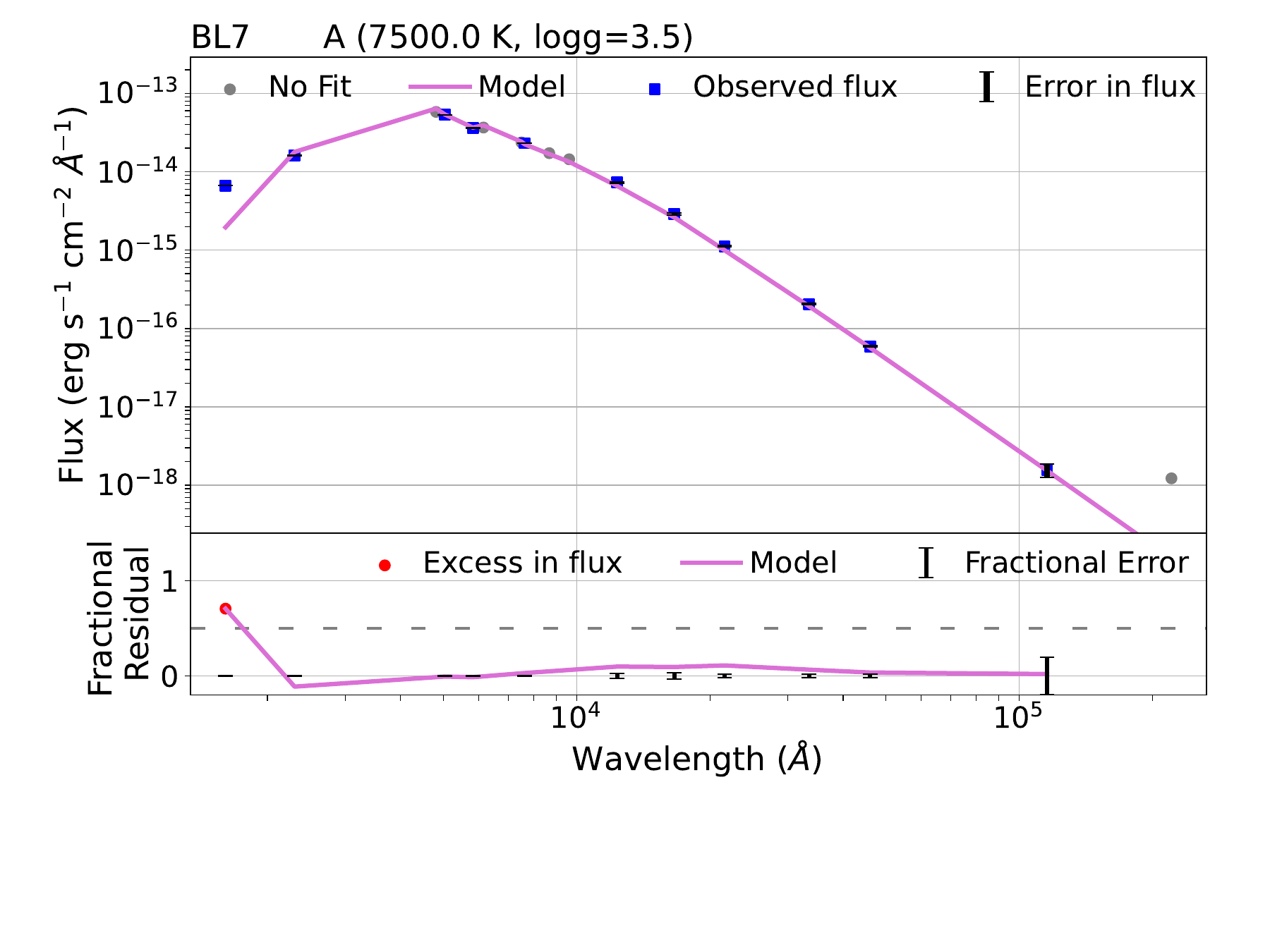} 
\includegraphics[width=0.45\textwidth]{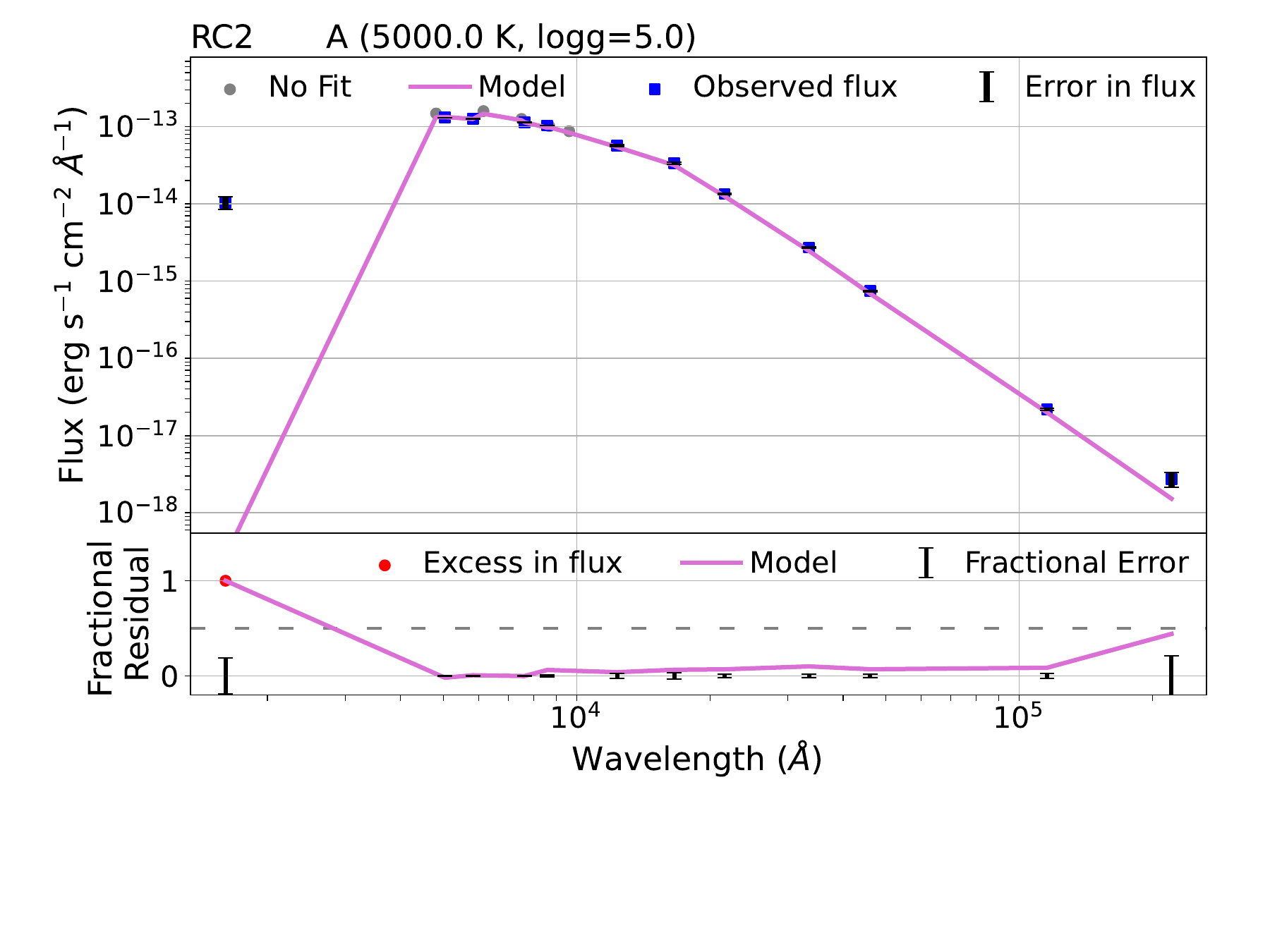} 
\caption{Sources showing excess in UV fluxes but fitted with the single component SEDs. The labels in this figure are the same as in Figure \ref{Fig.8}.}
\label{Fig.9}
\end{figure*}

\begin{figure*} 
\includegraphics[width=0.45\textwidth]{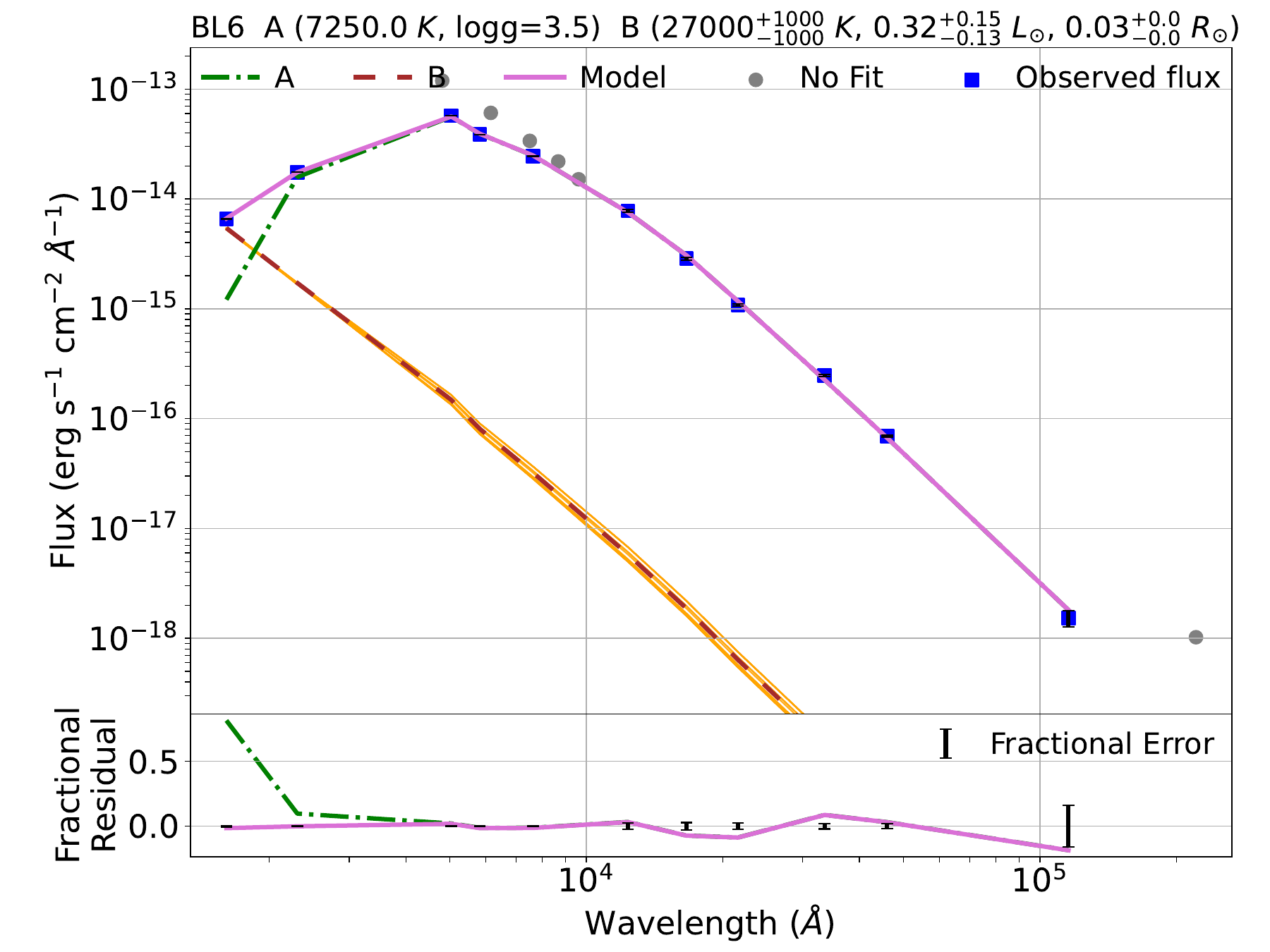} 
\includegraphics[width=0.45\textwidth]{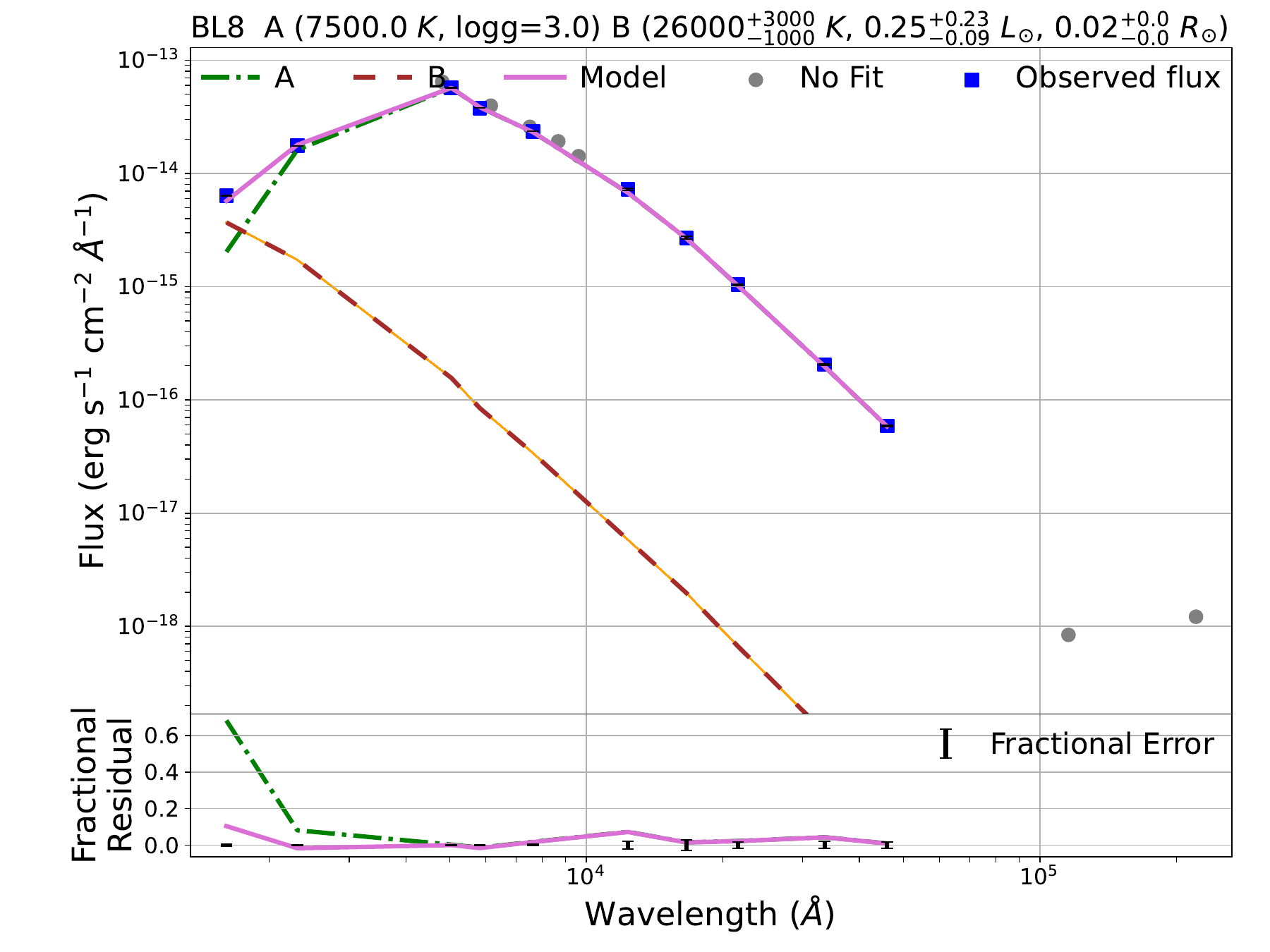} 
\includegraphics[width=0.45\textwidth]{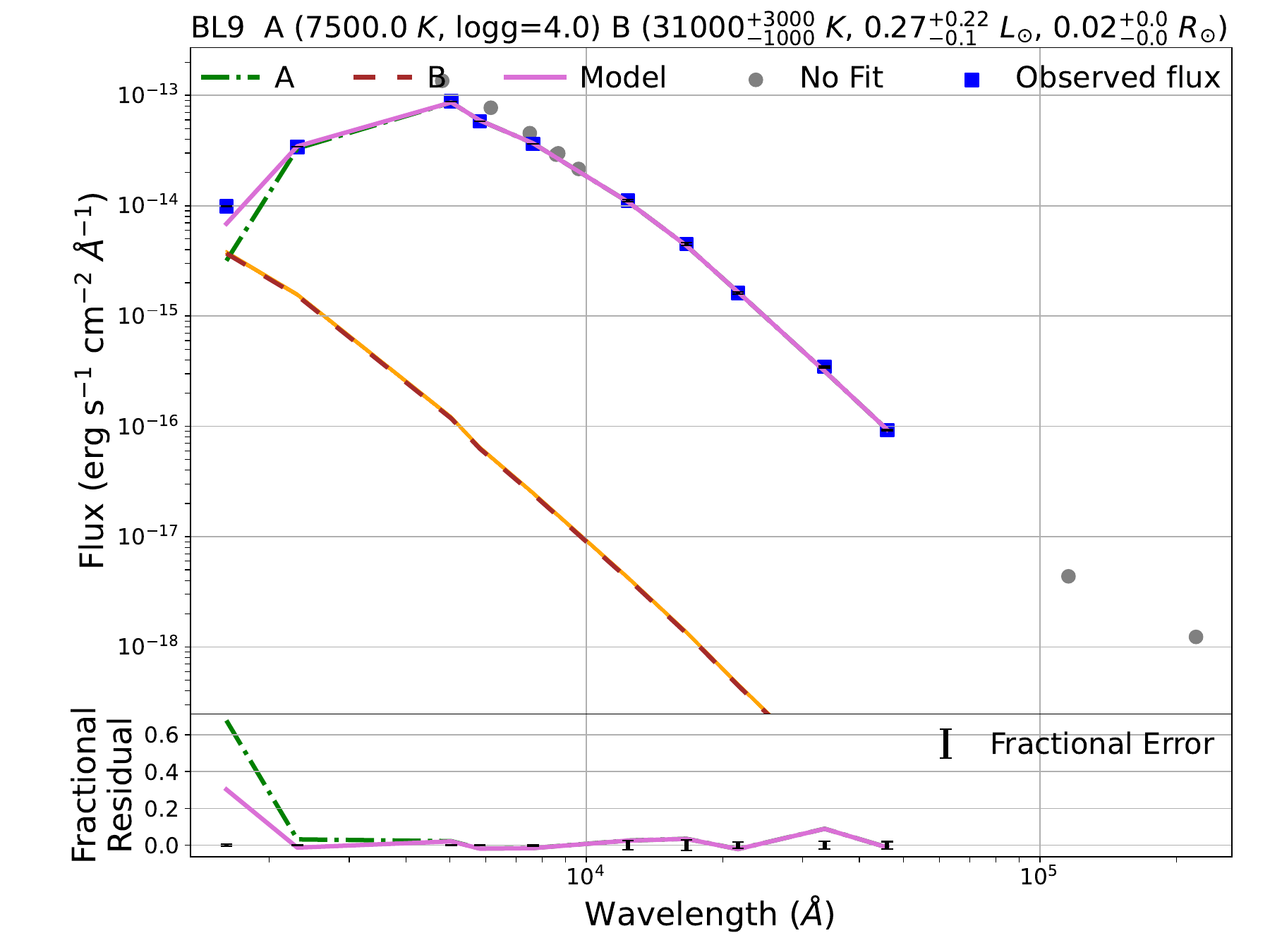} 
\includegraphics[width=0.45\textwidth]{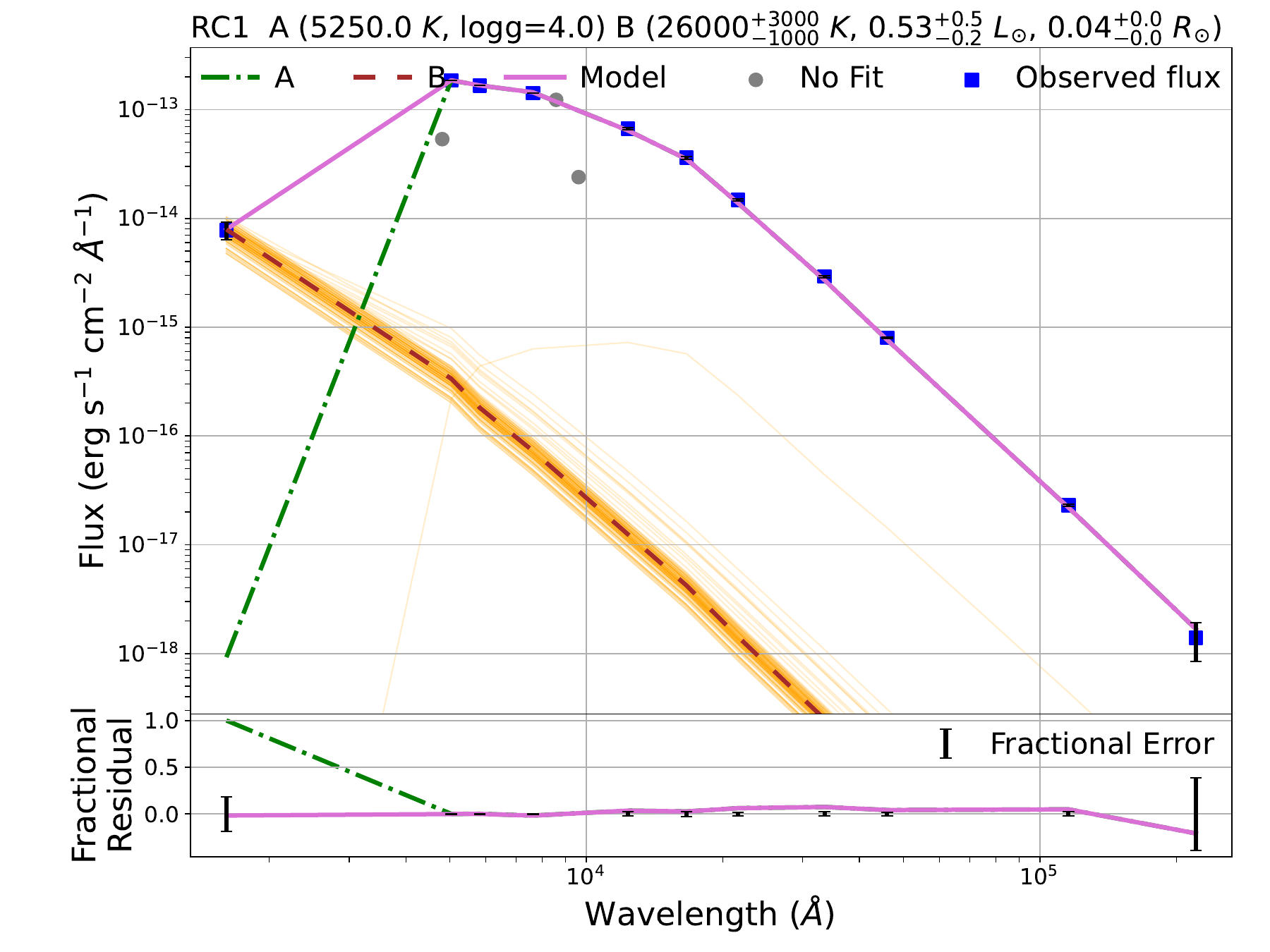}
\includegraphics[width=0.45\textwidth]{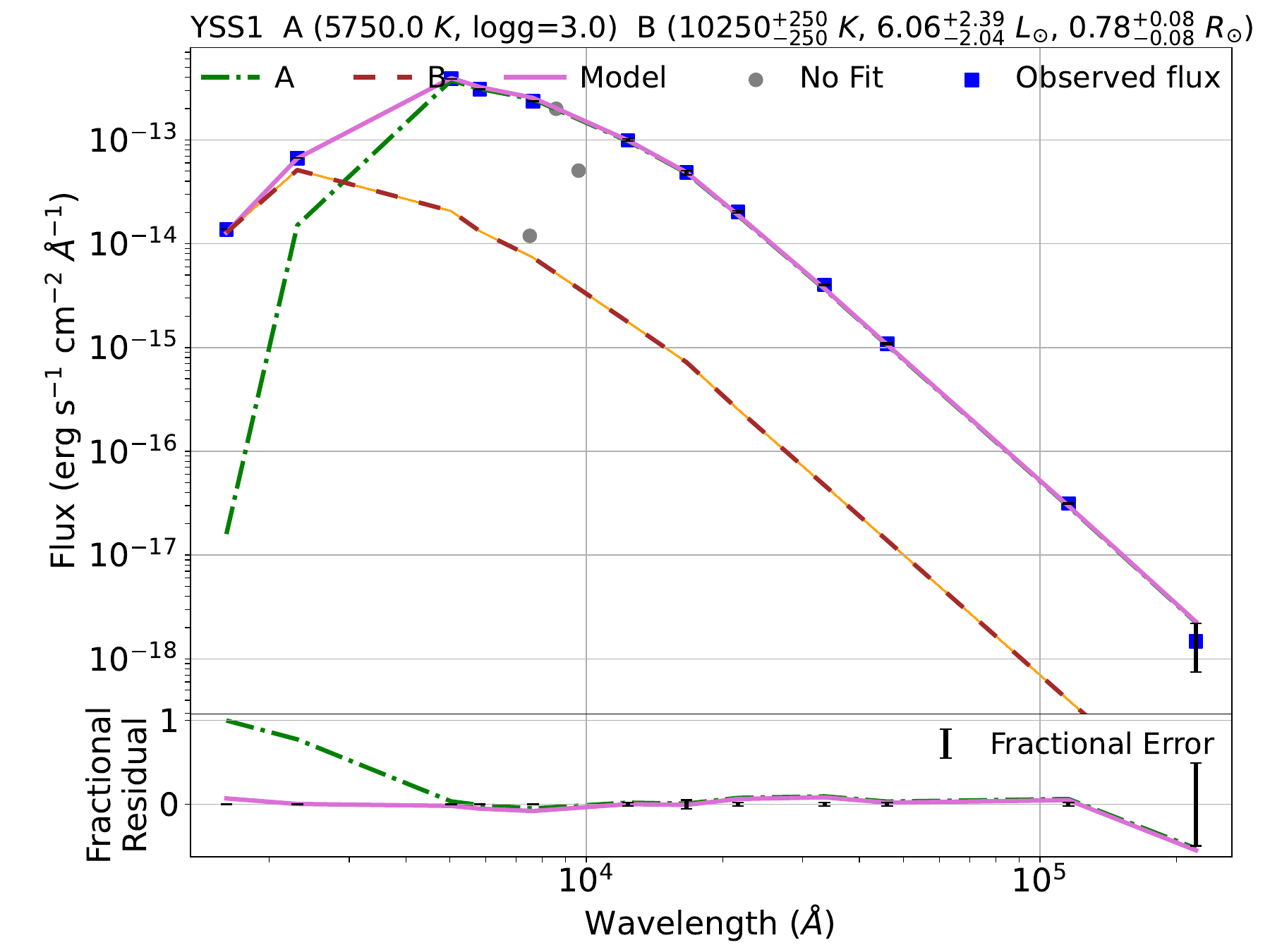} 
\includegraphics[width=0.45\textwidth]{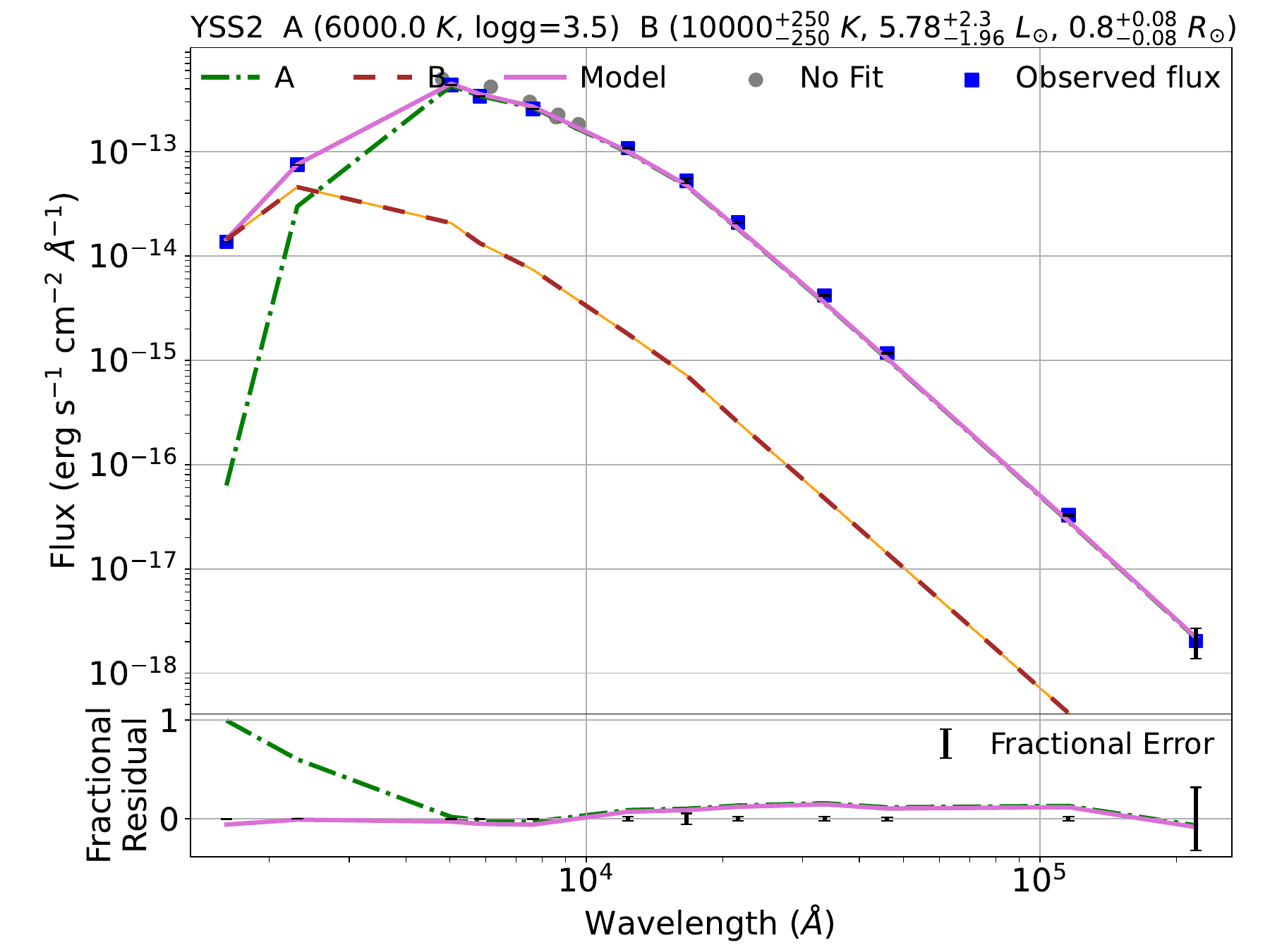}

\caption{The binary-component SEDs. The top panel displays the double component SED of each object, with blue data points representing extinction corrected flux values (labeled as \lq \lq Observed flux'') and black error bars showing flux errors. The green dashed line represents the cool (A) component fit (labeled as \lq \lq A''), while the brown dashed line represents the hot (B) component fit (labeled as \lq \lq B''), with orange curves representing iteration residuals. The composite fit is depicted as a purple curve (labeled as \lq \lq Model''), with grey data points denoting data points that were not included in the fits (labeled as \lq \lq No fit ''). The fractional residual for single (green) and composite (purple) fits is shown in the bottom panel. On the x-axis, black error bars represent fractional errors (labeled as \lq \lq Fractional error''). The cool and hot component parameters derived from the SED, as well as the estimated errors, are listed at the top of the plots.}
\label{Fig.10}
\end{figure*}

5) We observe that four BLs, two YSS, and two RC stars displayed an UV excess exceeding 50$\%$. Consequently, we attempt to fit double-component SEDs to these stars. Out of these eight stars, we could successfully fit the binary-component SEDs to six of them. In order to perform the binary-fit, we use a Python-based code \textsc{Binary SED Fitting}\footnote{https://github.com/jikrant3/Binary SED Fitting} developed by \cite{jadhav2021uocs}, which employs a $\chi^{2}$-minimization technique. However, for the one BL and one RC star, the models utilized in the aforementioned code fitted the UV data points by adopting their highest available temperatures. Consequently, we present the single-component SEDs for these two stars. The SEDs of sources showing UV excess but fitted with single-component SEDs are shown in Figure \ref{Fig.9}, whereas those fitted with binary-component SEDs are shown in Figure \ref{Fig.10} 

\section{Discussion} \label{Section 5}

The detection of exotic stars in clusters may be attributed to significant fraction of binaries in them. The open clusters NGC 188 and NGC 2682 have a large fraction of binaries ($\sim$ 0.51 and $\sim$ 0.47, respectively) and are found to have a significant number of BSS \citep{cordoni2023photometric}. This cluster also has a large fraction of binaries ($\sim$ 0.429) observed along the main sequence as reported by \cite{cordoni2023photometric}. Even despite its young age, as per our analysis, this cluster has 1 BSS, 11 BLs, and 2 YSS. The cluster also shows a clear eMSTO. The limited number of sources with the vsin\textit{i} information does not allow us to see any clear trend with the position of sources in the CMD, whereas the age spread (0.15 Gyr) found in the eMSTO sources is not negligible.
In several recent investigations, further intriguing sources, referred to as UV dim stars, have been identified in clusters such as NGC 1783, NGC 1850, NGC 2164, and NGC 1818 \citep{milone2022multiple, milone2023hubble, martocchia2023origin, d2023role} situated in the Large Magellanic Cloud. The authors concluded that these stars play a role in shaping the observed eMSTO phenomenon within these clusters.
These sources exhibit redder colors compared to the red sources within the eMSTO in UV CMDs. The presence of UV-dim stars in these young Magellanic cloud clusters is explained on the basis of high rotational velocities and the presence of an edge-on absorption dust ring. We checked for the presence of UV dim stars in NGC 6940. In the UVIT/F169M versus (UVIT/F169M-GALEX/NUV) CMD of this cluster we did not find any stars that were particularly redder compared to eMSTO. Therefore, we conclude that there are no UV dim stars in this cluster.

\subsection{Properties of BSS, BLs, YSS, and RC stars}

By following the aforementioned method to fit the SEDs of the BSS, BLs, YSS, and RC stars, we determined that the temperature of the BSS is 8500 K, while those of BLs vary between 7250 to 9750 K. Moreover, the YSS have temperatures 5750 and 6000 K, and that of RC stars are 5000 and 5250 K. The parameters of all the stars fitted with the single-component SEDs are listed in Table \ref{Table2}, whereas the parameters of stars fitted with the double-component SEDs are listed in Table \ref{Table3}.

On comparing the temperatures derived from SED fitting with temperatures based on low-resolution BP/RP spectra from \textit{Gaia} DR3 \citep{babusiaux2022gaia}, we found that the SED-based temperatures deviate by no more than 450 K. Additionally, we estimate the masses of BSS, BLs, YSS, and RC stars by comparing them with the zero-age main sequence (ZAMS) for an age of 8.5 Gyr. We found that the mass of BSS is equal to 2.09 M$_{\odot}$, the masses of BLs vary from 1.74 to 2.24 M$_{\odot}$, YSS vary from 2.53 to 2.80 M$_{\odot}$, and those of RC stars vary from 2.32 to 2.53 M$_{\odot}$. None of the above mentioned objects are found to be of variable nature in TESS data.

\begin{figure*}
\includegraphics[width=0.7\textwidth]{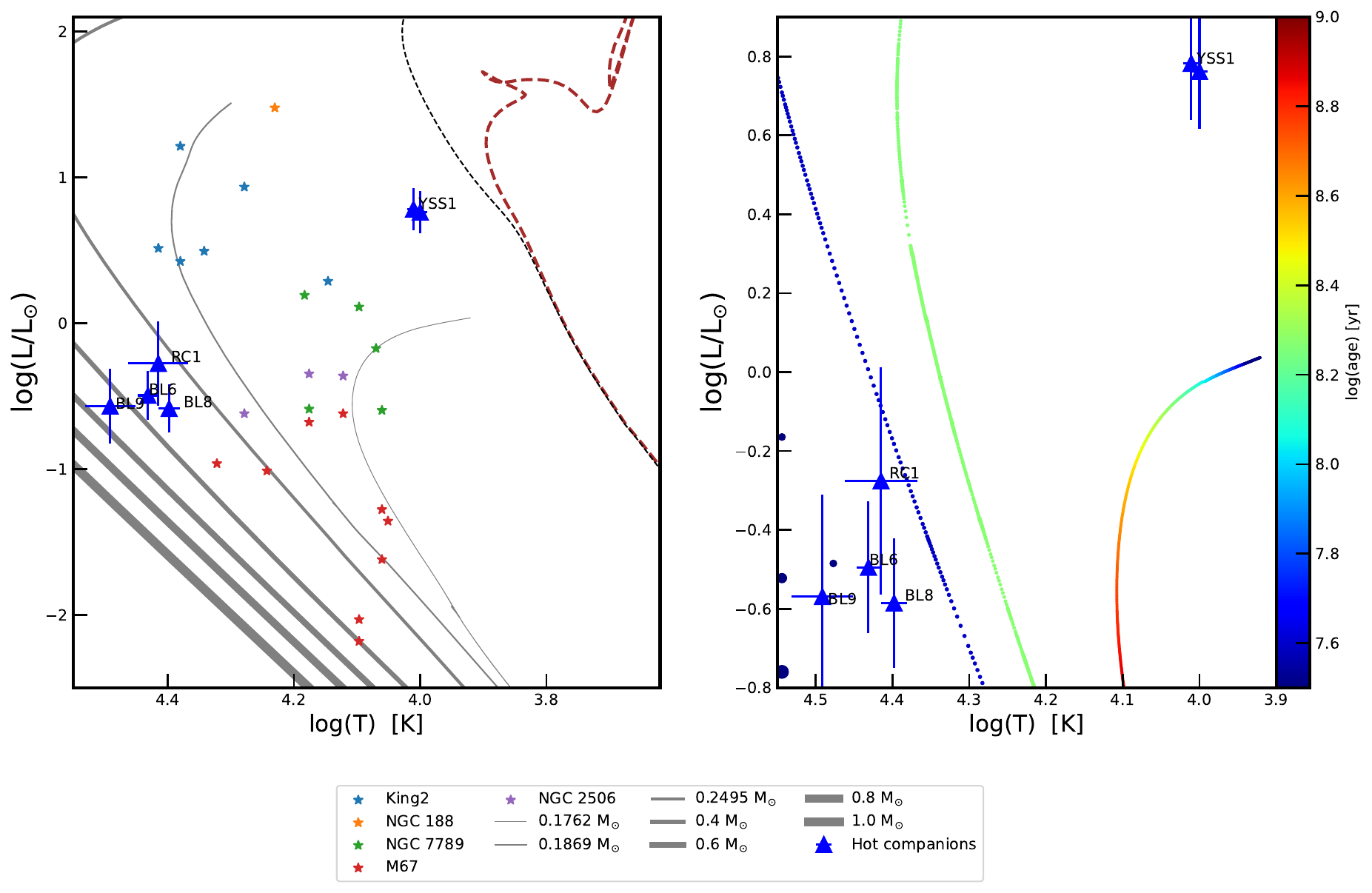} 
\caption{H-R diagram of NGC 6940. The left panel contains a PARSEC isochrone with an age of 1.0 Gyr and distance of 770 pc along with a ZAMS, which is constructed by stitching ZAMS of ages 8.0 Gyr, 8.5 Gyr, and 9.0 Gyr. We have shown the cooler and hot companions of NGC 6940 BLs, YSS, and RC stars along with the hot BSS companions of other open clusters. Moreover, the objects with single-component fits in NGC 6940 are shown with the mentioned labels. The WD cooling curves of different masses which are taken from \citet{althaus2013new} in case of ELM WDs, \citet{panei2007full} in the case of LM WDs, and \citet{tremblay2009spectroscopic} in the case of normal to high mass WDs are also shown. The right panel depicts the hot companions of the above mentioned objects lying on the WD cooling curves of different masses, indicating their approximate cooling ages.}
\label{Fig.11}
\end{figure*}

\subsection{The hot companions}

To attain a comprehensive understanding of the formation processes of the above mentioned exotic stars, it is imperative to examine the characteristics of their hot companions. In pursuit of this goal, we construct the Hertzsprung-Russell (H-R) diagram, as depicted in Figure \ref{Fig.11}. We have shown the objects fitted with single-component fits, cool and hot companions of binary-component fits, with the symbols as labeled in the figure. Below we describe the nature of hot companions in detail.

\subsubsection{The hot companions of BLs}

We had classified eleven sources observed in UVIT as BL candidates due to their higher rotational velocities, of which two BL candidates had nearby sources. Out of nine sources, we successfully fit the binary-component SEDs to three of them and single-component SEDs of five BL candidates. In these systems, the hot companions of BL6 and BL8 fall between the WD cooling curves of masses 0.2 -- 0.4 M$_{\odot}$, classifying them as LM WD \citep{panei2007full}, as shown in Figure \ref{Fig.10}. LM WD cannot form from a single star evolution in the Hubble time, hence, this is evidence of MT via Case-A/Case-B. On the other hand, the hot companion of BL9 has an estimated mass $\sim$0.5 M$_{\odot}$, implying that it is a normal mass WD, suggesting either Case-C MT or merger as the possible formation channel. The evidence of MT in these 3 BL candidates confirms that these three are genuine BLs. The MT also explains the enhanced vsin\textit{i} values in these three objects.

Regarding the remaining BL candidate exhibiting UV excess, the excess might be indicative of a hot companion. This UV excess could also be due to other factors, such as chromospheric activities or the presence of hot spots. The discovery of these three BLs makes this cluster the third open cluster \citep{leiner2019blue, mcclure2021search} with the presence of BLs. Spectroscopic observations of these BLs will help us confirm their formation scenario by abundance analysis.

 \subsubsection{The hot companions of YSS}

The positioning of the hot companions of both the YSS on the H-R diagram indicates that they are subdwarf B (sdB) stars. These stars are considered to be objects that burn helium in their cores while having very thin envelopes containing hydrogen \citep{sargent1968quantitative}. Recent observational surveys have revealed that a significant number of these stars are part of binary systems. There are three primary pathways in binary evolution that can result in the formation of sdB stars. The first is the common-envelope ejection channel, where sdB stars are formed in binaries with short orbital periods ranging from 0.1 to 10 days. These stars have thin hydrogen-rich envelopes. Alternatively, the stable Roche lobe overflow channel, assuming that most of the transferred mass is lost, leads to the production of sdB stars. However, these stars have longer orbital periods ranging from 400 to 1500 days and possess relatively thicker hydrogen-rich envelopes. The third pathway, known as the double helium WDs merger channel, results in the formation of single sdB stars. These stars have extremely thin hydrogen-rich envelopes \citep{han2002origin}. Since, both sdB stars are found as the hot companions of YSS, we suggest that the sdB progenitors transferred the mass to their companions, which became YSS.

\subsubsection{The hot companions of RC stars}

As mentioned above, both the RC stars showed the excess in UV fluxes indicating the presence of hot companions. However, we could fit the binary-component SED to only one of them. The location of the hot companion of RC1 in the H-R diagram suggests it to be a LM WD of mass $\sim$0.3 M$_{\odot}$. This implies that the RC star has formed via Case-A/Case-B MT. On the other hand, a significant UV excess in the flux of RC2 suggests that the hot companion may be responsible for this, however we are unable to successfully fit the second component SED to this star.
\section{Summary and conclusions} \label{Section 6}

The main findings of this paper can be summarized as follows:

\begin{itemize}

\item We use a machine learning algorithm, ML-MOC, on \textit{Gaia} DR3 data to successfully identify 492 cluster members, including one BSS, eleven BL candidates, two YSS, and two RC stars. We use the \textit{AstroSat}/UVIT data in F169M filter to characterize these objects. \\

\item The King's profile fitting to the radial density of members provides us the core radius of the cluster as 9.8$\arcmin$ and the tidal radius of the cluster as 37.5$\arcmin$. \\

\item The massive single stars show highest degree of segregation followed by the equal-mass binary main sequence populations and then the low-mass single stars. This observation provides evidence that the cluster is indeed mass-segregated, indicating that dynamic evolution has occurred within the cluster. \\

\item We report the presence of eMSTO feature in this cluster for the first time in the literature. The median age that fits the eMSTO stars is found to be 0.7 Gyr with 0.15 Gyr as the 1-$\sigma$ error. This age spread is large enough to cause the eMSTO feature. Moreover, we found that the variations in the vsin\textit{i} when available (12$\%$ of the MSTO stars) are significant, varying from 10 km/s to 150 km/s. This suggests that rotation may contribute to the observed eMSTO phenomenon. In order to investigate the real cause of the eMSTO it is crucial to obtain measurements of the rotational velocities of large number of stars along the main sequence. Furthermore, the effect of circumstellar dust can be examined to conclude about the possible reason for the eMSTO.\\

\item Out of sixteen exotic stars (BSS, BLs, YSS, and RC stars) detected in the UVIT, we construct the multi-wavelength SED of fourteen of them. Two BLs had a neighbouring source within 3$\arcsec$, and hence their SEDs are not constructed. 

\item We fit the BSS with the single-component SED. The absence of excess in UV fluxes suggest that a cool WD may be present as the companion of this star, which cannot be detected by UVIT, or that this BSS have formed as a result of merger.

\item Out of eleven BL candidates, we study nine of them as the remaining two have nearby sources within 3$\arcsec$. Out of these nine stars, three are fitted with the binary-component SEDs. Two of them are likely to have a LM WD and the third one has a normal mass WD as the hot companion. The discovery of LM WDs as the hot companion suggests that the BLs are formed via Case-B MT, whereas the discovery of normal-mass WD indicates either Case-C or merger as the possible formal channels. Five BL candidates are fitted with the single-component SEDs, out of which three have high rotational velocities (vsin\textit{i} $>$ 100 km/s). The lack of excess in UV fluxes suggests that they may harbour a cool WD or be formed via merger. The remaining one BL candidate has UV excess but could not be fitted with the binary-component SEDs using any of the models. Therefore, we conclude that out of the nine BL candidates, at least three are confirmed BLs.\\ 

\item We find two sdB stars as likely hot companions of YSS, and one normal mass WD as hot companion of RC star. The discovery of sdB stars as the hot companions also indicates the MT formation channel. Overall, $\sim$42$\%$ of the objects studied here are likely to have formed through the MT and/or merger channels. \\

\item The two objects (one BL and one RC star) exhibiting excess in UV fluxes, but not fitting the binary-component SEDs, could still be binary systems. We need data in multiple UV filters and additional spectroscopic observations to constrain the SEDs and learn the true nature of these objects. 

\end{itemize}

\section{Acknowledgements}
We thank the anonymous referee for the constructive feedback and suggestions that has helped in improving the quality of the paper. The authors would like to thank Manan Agarwal for developing the ML-MOC algorithm, using which the cluster membership has been determined for this work. This work uses the data from UVIT onboard AstroSat mission of Indian Space Research Organisation (ISRO). UVIT is a collaborative project between Indian Institute of Astrophysics (IIA), Bengaluru, The Indian-University Centre for Astronomy and Astrophysics (IUCAA), Pune, Tata Institute of Fundamental Research (TIFR), Mumbai, several centres of Indian Space Research Organisation (ISRO), and Canadian Space Agency (CSA). This paper also includes data collected by the TESS mission, which are publicly available from the Mikulski Archive
for Space Telescopes (MAST). Funding for the TESS mission is provided by NASA’s Science Mission Directorate. This publication also makes use of VOSA, developed under the Spanish Virtual Observatory project supported by the Spanish MINECO through grant AyA2017-84089. 

\section{Data availability}
The data underlying this article are publicly available at \url{https://astrobrowse.issdc.gov.in/astro_archive/archive/Home.jsp} The derived data generated in this research will be shared on reasonable request to the corresponding author.

\bibliography{references}
\bibliographystyle{MNRAS}

\begin{table*}
\centering
\caption{Coordinates of all the sources in Columns 2 and 3, UVIT F169M flux in Columns 4, GALEX NUV flux in Column 5, \textit{Gaia} DR3 and PANSTARRS fluxes in Columns 6--13, 2MASS J, H, and Ks fluxes in Columns 14--16, and WISE W1, W2, W3, and W4 fluxes in Columns 17--20. All flux values are extinction corrected and listed in the unit of erg s$^{-1}$ cm$^{-2}$\AA$^{-1}$.}

\begin{tabular}{cccccccccccc}
\hline
Name&RA&DEC&UVIT.F169M$\pm$err&GALEX.NUV$\pm$err&\\
PS1.g$\pm$err&GAIA3.Gbp$\pm$err&GAIA3.G$\pm$err&PS1.r$\pm$err&PS1.i$\pm$err&\\
GAIA3.Grp$\pm$err&PS1.z$\pm$err&PS1.y$\pm$err&2MASS.J$\pm$err&2MASS.H$\pm$err&\\
2MASS.Ks$\pm$err&WISE.W1$\pm$err&WISE.W2$\pm$err&WISE.W3$\pm$err&WISE.W4$\pm$err \\
\hline
\hline
\\
BSS1&308.79705&28.40223&2.667e-14$\pm$5.549e-17&1.138e-13$\pm$1.216e-17&\\
3.653e-13$\pm$0.000e+00&1.796e-13$\pm$4.779e-16&1.064e-13$\pm$2.721e-16&1.584e-13$\pm$0.000e+00&8.060e-14$\pm$0.000e+00&\\
5.784e-14$\pm$2.062e-16&5.013e-14$\pm$0.000e+00&3.686e-14$\pm$0.000e+00&1.609e-14$\pm$4.001e-16&5.291e-15$\pm$1.803e-16&\\
2.073e-15$\pm$4.774e-17&4.370e-16$\pm$8.854e-18&1.206e-16$\pm$2.221e-18&3.309e-18$\pm$2.804e-19&1.164e-18$\pm$0.000e+00\\

\hline

BL1&308.75408&28.28973&1.671e-14$\pm$3.625e-17&8.216e-14$\pm$1.210e-17&\\
2.086e-13$\pm$0.000e+00&1.595e-13$\pm$4.151e-16&1.008e-13$\pm$2.585e-16&1.176e-13$\pm$0.000e+00&7.244e-14$\pm$0.000e+00\\
5.989e-14$\pm$2.092e-16&5.144e-14$\pm$0.000e+00&3.622e-14$\pm$0.000e+00&1.832e-14$\pm$3.881e-16&6.606e-15$\pm$1.765e-16\\
2.481e-15$\pm$4.114e-17&4.818e-16$\pm$9.318e-18&1.364e-16$\pm$2.387e-18&4.846e-18$\pm$2.990e-19&2.262e-18$\pm$0.000e+00\\

\hline

\hline

BL4&309.03813&28.43106&1.037e-14$\pm$3.407e-17&4.913e-14$\pm$1.210e-17\\
2.108e-13$\pm$0.000e+00&1.413e-13$\pm$3.664e-16&9.150e-14$\pm$2.328e-16&1.109e-13$\pm$0.000e+00&6.456e-14$\pm$0.000e+00\\
5.567e-14$\pm$1.941e-16&4.484e-14$\pm$0.000e+00&3.424e-14$\pm$0.000e+00&1.655e-14$\pm$3.354e-16&6.081e-15$\pm$1.456e-16&\\
2.316e-15$\pm$4.692e-17&4.509e-16$\pm$9.551e-18&1.271e-16$\pm$2.575e-18&3.569e-18$\pm$2.761e-19&1.275e-18$\pm$0.000e+00\\

\hline

BL5&308.64241&28.36945&1.053e-14$\pm$2.753e-17&3.570e-14$\pm$1.205e-17\\
7.499e-14$\pm$0.000e+00&6.714e-14$\pm$1.756e-16&4.264e-14$\pm$1.084e-16&4.349e-14$\pm$0.000e+00&2.726e-14$\pm$0.000e+00\\
2.517e-14$\pm$8.795e-17&1.974e-14$\pm$0.000e+00&1.469e-14$\pm$3.950e-17&7.213e-15$\pm$1.727e-16&2.601e-15$\pm$6.948e-17\\
9.942e-16$\pm$1.648e-17&1.932e-16$\pm$4.093e-18&5.511e-17$\pm$1.117e-18&1.882e-18$\pm$2.669e-19&1.461e-18$\pm$0.000e+00\\

\hline 

BL6&308.63418&28.49372&6.549e-15$\pm$3.440e-17&1.749e-14$\pm$1.194e-17\\
1.194e-13$\pm$0.000e+00&5.734e-14$\pm$1.488e-16&3.861e-14$\pm$9.812e-17&6.076e-14$\pm$0.000e+00&3.379e-14$\pm$0.000e+00\\
2.459e-14$\pm$8.578e-17&2.194e-14$\pm$0.000e+00&1.510e-14$\pm$3.488e-17&7.793e-15$\pm$1.866e-16&2.860e-15$\pm$7.902e-17&\\
1.077e-15$\pm$2.480e-17&2.469e-16$\pm$5.229e-18&6.893e-17$\pm$1.333e-18&1.524e-18$\pm$2.470e-19&1.022e-18$\pm$0.000e+00\\

\hline 

BL7&308.58822&28.57094&6.657e-15$\pm$3.309e-17&1.610e-14$\pm$1.194e-17\\
5.800e-14$\pm$0.000e+00&5.339e-14$\pm$1.391e-16&3.613e-14$\pm$9.185e-17&3.665e-14$\pm$0.000e+00&2.366e-14$\pm$0.000e+00\\
2.317e-14$\pm$8.096e-17&1.725e-14$\pm$0.000e+00&1.442e-14$\pm$3.831e-17&7.286e-15$\pm$1.812e-16&2.878e-15$\pm$9.014e-17\\
1.118e-15$\pm$2.059e-17&2.048e-16$\pm$4.338e-18&5.878e-17$\pm$1.191e-18&1.564e-18$\pm$3.068e-19&1.223e-18$\pm$0.000e+00\\

\hline

BL8&308.78129&28.34795&6.387e-15$\pm$2.629e-17&1.760e-14$\pm$1.194e-17\\
6.472e-14$\pm$0.000e+00&5.709e-14$\pm$1.484e-16&3.787e-14$\pm$9.626e-17&3.970e-14$\pm$0.000e+00&2.582e-14$\pm$0.000e+00\\
2.361e-14$\pm$8.240e-17&1.924e-14$\pm$0.000e+00&1.422e-14$\pm$2.356e-17&7.280e-15$\pm$1.542e-16&2.706e-15$\pm$7.727e-17\\
1.041e-15$\pm$1.726e-17&2.055e-16$\pm$3.975e-18&5.889e-17$\pm$1.030e-18&8.413e-19$\pm$0.000e+00&1.216e-18$\pm$0.000e+00\\

\hline

BL9&308.60522&28.27182&9.845e-15$\pm$4.632e-17&3.414e-14$\pm$1.205e-17\\
1.357e-13$\pm$0.000e+00&8.826e-14$\pm$2.292e-16&5.828e-14$\pm$1.481e-16&7.735e-14$\pm$0.000e+00&4.549e-14$\pm$0.000e+00\\
3.637e-14$\pm$1.269e-16&2.987e-14$\pm$0.000e+00&2.156e-14$\pm$2.552e-17&1.113e-14$\pm$2.768e-16&4.495e-15$\pm$1.242e-16\\
1.617e-15$\pm$2.681e-17&3.474e-16$\pm$7.360e-18&9.256e-17$\pm$1.790e-18&4.371e-18$\pm$3.220e-19&1.234e-18$\pm$0.000e+00\\

\hline

BL10&308.83537&28.23503&1.247e-14$\pm$3.400e-17&5.341e-14$\pm$1.210e-17&\\
1.285e-13$\pm$0.000e+00&9.987e-14$\pm$2.638e-16&6.303e-14$\pm$1.608e-16&7.165e-14$\pm$0.000e+00&4.349e-14$\pm$0.000e+00\\
3.701e-14$\pm$1.298e-16&3.082e-14$\pm$0.000e+00&2.167e-14$\pm$5.445e-17&1.100e-14$\pm$2.127e-16&4.029e-15$\pm$1.076e-16\\
1.520e-15$\pm$2.520e-17&2.879e-16$\pm$6.099e-18&8.196e-17$\pm$1.661e-18&2.349e-18$\pm$2.618e-19&1.148e-18$\pm$0.000e+00\\

\hline

YSS1&308.89017&28.27984&1.375e-14$\pm$3.871e-17&6.662e-14$\pm$1.210e-17\\
-&3.876e-13$\pm$1.002e-15&3.070e-13$\pm$7.805e-16&-&1.190e-14$\pm$1.307e-17\\
2.357e-13$\pm$8.211e-16&-&5.059e-14$\pm$4.660e-17&9.929e-14$\pm$2.378e-15&4.844e-14$\pm$2.454e-15\\
2.024e-14$\pm$3.915e-16&4.015e-15$\pm$8.505e-17&1.088e-15$\pm$1.905e-17&3.137e-17$\pm$6.068e-19&1.473e-18
$\pm$7.298e-19\\
\hline

YSS2&308.59549&28.36145&1.362e-14$\pm$3.535e-17&7.495e-14$\pm$1.210e-17\\
4.973e-13$\pm$0.000e+00&4.366e-13$\pm$1.126e-15&3.388e-13$\pm$8.609e-16&4.192e-13$\pm$0.000e+00&3.000e-13$\pm$0.000e+00\\
2.557e-13$\pm$8.910e-16&2.258e-13$\pm$0.000e+00&1.832e-13$\pm$0.000e+00&1.082e-13$\pm$2.092e-15&5.233e-14$\pm$2.844e-15\\
2.098e-14$\pm$4.445e-16&4.177e-15$\pm$8.848e-17&1.163e-15$\pm$2.143e-17&3.261e-17$\pm$6.608e-19&2.029e-18$\pm$6.467e-19\\

\hline

RC1&308.56125&28.37103&7.789e-15$\pm$1.428e-15&-&\\
5.354e-14$\pm$1.285e-15&1.858e-13$\pm$4.829e-16&-&-\\
1.668e-13$\pm$4.239e-16&1.419e-13$\pm$4.949e-16&2.395e-14$\pm$2.394e-17&6.664e-14$\pm$1.534e-15&3.610e-14$\pm$9.976e-16\\
1.475e-14$\pm$2.716e-16&2.922e-15$\pm$6.189e-17&7.988e-16$\pm$1.471e-17&2.309e-17$\pm$5.528e-19&1.389e-18	$\pm$5.397e-19\\	

\hline

RC2&308.59290&28.56243&1.033e-14$\pm$1.923e-15&-\\
1.480e-13$\pm$0.000e+00&1.307e-13$\pm$3.424e-16&1.265e-13$\pm$3.227e-16&1.588e-13$\pm$0.000e+00&1.258e-13$\pm$0.000e+00\\
1.139e-13$\pm$3.981e-16&1.015e-13$\pm$0.000e+00	&8.712e-14$\pm$0.000e+00&5.656e-14$\pm$1.407e-15&3.379e-14$\pm$1.058e-15\\
1.346e-14$\pm$2.232e-16&2.724e-15$\pm$5.771e-17&7.413e-16$\pm$1.434e-17&2.176e-17$\pm$6.014e-19&2.730e-18$\pm$5.834e-19\\

\hline

BL11&308.73803&28.22449&1.165e-13$\pm$8.215e-17&1.332e-13$\pm$1.216e-17\\
1.945e-13$\pm$0.000e+00&1.122e-13$\pm$2.927e-16&6.468e-14$\pm$1.646e-16&8.364e-14$\pm$0.000e+00&4.199e-14$\pm$0.000e+00\\
3.289e-14$\pm$1.149e-16&2.545e-14$\pm$0.000e+00&1.818e-14$\pm$5.349e-17&8.056e-15$\pm$1.929e-16&2.810e-15$\pm$8.024e-17\\
1.017e-15$\pm$1.874e-17&1.957e-16$\pm$4.146e-18&5.366e-17$\pm$1.038e-18&1.621e-18$\pm$2.792e-19&1.084e-18$\pm$0.000e+00\\

\hline

\label{Table1}
\end{tabular}

\end{table*}

\begin{table*}
\caption{The best-fit parameters of all stars fitted with the single-component SEDs. For each of them, we have listed  luminosity, temperature, and radius in Columns 2 $-$ 4, the reduced $\chi^{2}_{r}$ values in Column 5, the scaling factor in Column 6, the number of data points used to fit the SED is given in Column 7, and the values of vgf$_{b}$ parameter in Column 8.} 
\adjustbox{max width=\textwidth}{
\begin{tabular}{ccccccccccc}
\hline
\\
Name&Luminosity&T$\mathrm{_{eff}}$&Radius&$\chi^{2}_{r}$&Scaling factor&N$_{fit}$&$vgf_{b}$\\
~&~&[L$_\odot$]&[K]&[R$_\odot$]&&~&~&
\\
\hline 
\\
BSS1&32.58$\pm$0.02&8500$\pm$125&2.63$\pm$0.01&4.21&3.53&11&0.27\\

BL1&30.77$\pm$0.02&8000$\pm$125&2.88$\pm$0.00&27.44&4.238e-21&13&0.928\\

BL4&26.63$\pm$0.01&7750$\pm$125&2.86$\pm$0.09&27.08&4.171e-21&13&0.194\\

BL5&24.19$\pm$0.02&7500$\pm$125&2.91$\pm$0.09&31.31&4.319e-21&13&0.08\\

BL7&10.53$\pm$0.01&7500$\pm$125&1.92$\pm$0.06&25.98&1.87e-21&12&0.535\\

BL10&19.15$\pm$0.01&8000$\pm$125&2.27$\pm$0.01&24.15&2.640e-21&13&0.118\\ 	

BL11&24.27$\pm$0.01&9750$\pm$125&1.72$\pm$0.01&6.31&1.51e-21&11&0.23\\

RC2&40.71$\pm$8.24&5000$\pm$125&8.45$\pm$0.08&25.29&3.635e-20&13&0.997\\
\hline

\label{Table2}
\end{tabular}
}
\end{table*}

\begin{table*}
\caption{The best-fit parameters of all stars fitted with the double-component SEDs. For each of them, whether cooler (A) or hot (B) companion in Column 2, model used for fitting Column 3, luminosity, temperature, and radius in Columns 4 $-$ 6, the reduced $\chi^{2}_{r}$ values in Column 7 (the $\chi^{2}_{r}$ values of the single fits are given in the brackets), scaling factor in Column 8, number of data points used to fit the SEDs is given in Column 9, and the values of vgf$_{b}$ parameter in Column 10 (the vgf$_{b}$ values of the single fits are given in the brackets).}

\adjustbox{max width=\textwidth}{
\begin{tabular}{cccccccccccc}
\hline
\\
Name&Component&Model&Luminosity&T$\mathrm{_{eff}}$&Radius&$\chi^{2}_{r}$&Scaling factor&N$_{fit}$&vgf$_{b}$\\
~&~&~&[L$_\odot$]&[K]&[R$_\odot$]&&~&~&
\\
\hline 
\\

BL6&A&Kurucz&11.11$\pm$0.01&7250$\pm$125&2.11$\pm$0.07&12.85 (20.99)&2.276e-21&12&0.32(0.52)\\
&B&Koester&0.32$^{+0.15}_{-0.13}$&27000$\pm$1000&0.03$\pm$0.00&&2.642e-25&&\\

BL8&A&Kurucz&10.99$\pm$0.01&7500$\pm$125&1.96$\pm$0.06&9.57 (13.42)&1.963e-21&12&0.15 (0.58)\\ 	 	
&B&Kurucz&0.25$^{+0.23}_{0.09}$&26000$\pm$3000&0.02$\pm$0.00&&1.0424e-24&&\\

BL9&A&Kurucz&17.26$\pm$0.011&7500$\pm$125&2.46$\pm$0.08&28.95 (747.66)&3.082e-21&13&2.41 (32.48)\\
&B&Kurucz&0.27$^{+0.22}_{-0.1}$&31000$\pm$1000&0.02$\pm$0.00&&1.672e-25&&\\

YSS1&A&Kurucz&89.89$\pm$0.20&5750$\pm$125&9.55$\pm$0.41& 54.62 (213.70)&4.637e-20&13&0.38 (0.43) \\ 	
&B&Koester&6.06$^{+2.39}_{-2.04}$&10250$\pm$250&0.78$\pm$0.08&&3.099e-22&&\\

YSS2&A&Kurucz&99.49$\pm$0.221&6000$\pm$125&9.23$\pm$0.38&429.78 (460.16)&4.289e-20&13&0.94 (1.17)\\
&B&Koester&5.78$^{+2.3}_{-1.96}$&10000$\pm$250&0.8$\pm$0.08&&3.260e-22&&\\

RC1&A&Kurucz&51.36$\pm$10.37&5250$\pm$125&8.64$\pm$0.86&8.60 (22.64)&3.80e-20&12&0.23 (3.51)\\
&B&Kurucz&0.53$^{+0.5}_{-0.2}$&26000$^{+3000}_{-1000}$&0.04$\pm$0.00&&4.198e-22&&\\
\hline  
\label{Table3}
\end{tabular}
}
\end{table*}
\bsp	
\label{lastpage}
\end{document}